\documentclass[codesnippet,nojss,shortnames]{jss}
\usepackage{thumbpdf,lmodern}
\graphicspath{{Figures/}}

\usepackage{amsmath,amssymb,amsfonts}
\usepackage{tikz}

\newcommand{\dn}{\textrm{d}}       
\newcommand{\dd}{\, \textrm{d}}       
\newcommand{\dint}{\int \!}       

\definecolor{plotColors0}{HTML}{1B9E77}
\definecolor{plotColors1}{HTML}{D95F02}
\definecolor{plotColors2}{HTML}{7570B3}
\definecolor{plotColors3}{HTML}{E7298A}
\definecolor{plotColors4}{HTML}{66A61E}
\definecolor{plotColors5}{HTML}{E6AB02}
\definecolor{bg}{rgb}{0.9,0.9,0.9}

\author{Johan Dahlin\\Link\"{o}ping University
  \And Thomas
  B.\ Sch\"{o}n\\Uppsala University}

\title{Getting Started with Particle Metropolis-Hastings \\ for
  Inference in Nonlinear Dynamical Models}

\Plainauthor{Johan Dahlin,  Thomas B.\ Sch\"{o}n}

\Plaintitle{Getting Started with Particle Metropolis-Hastings for Inference in Nonlinear Dynamical Models}

\Shorttitle{Getting Started with PMH for Inference in Nonlinear Dynamical Models}

\Abstract{This tutorial provides a gentle introduction to the particle
  Metropolis-Hastings (PMH) algorithm for parameter inference in
  nonlinear state-space models together with a software implementation
  in the statistical programming language \proglang{R}. We employ a
  step-by-step approach to develop an implementation of the PMH
  algorithm (and the particle filter within) together with the
  reader. This final implementation is also available as the package
  \pkg{pmhtutorial} from the Comprehensive \proglang{R} Archive
  Network (CRAN) repository. Throughout the tutorial, we provide some
  intuition as to how the algorithm operates and discuss some
  solutions to problems that might occur in practice. To illustrate
  the use of PMH, we consider parameter inference in a linear Gaussian
  state-space model with synthetic data and a nonlinear stochastic
  volatility model with real-world data.  }


\Plainkeywords{Bayesian inference, state-space models, particle filtering, particle Markov chain Monte Carlo, stochastic volatility model}

\Address{
  Johan Dahlin\\
  Department of Computer and Information Science\\
  Link\"{o}ping University\\
  SE-581 83 Link\"{o}ping, Sweden\\
  E-mail: \email{uni@johandahlin.com}\\
  URL: \url{http://www.johandahlin.com/} \\ \\

  Thomas B.\ Sch\"{o}n\\
  Department of Information Technology\\
  Uppsala University\\
  Box 337 \\
  SE-751 05 Uppsala, Sweden \\
  E-mail: \email{thomas.schon@it.uu.se}\\
  URL: \url{http://user.it.uu.se/~thosc112/}
}

\begin{document}

\section{Introduction}
\label{sec:introduction}
We are concerned with Bayesian parameter inference in nonlinear
state-space models (SSMs). This is an important problem as SSMs are
ubiquitous in, e.g., automatic control \citep{Ljung1999}, econometrics
\citep{DurbinKoopman2012}, finance \citep{Tsay2005} and many other
fields. An overview of some concrete applications of state-space
modeling is given by \cite{Langrock2011}.

The major problem with Bayesian parameter and state inference in SSMs
is that it is an analytically intractable problem, which cannot be
solved in closed-form. Hence, approximations are required and these
typically fall into two categories: analytical approximations and
statistical simulations. The focus of this tutorial is on a member of
the latter category known as particle Metropolis-Hastings (PMH;
\citealp{AndrieuDoucetHolenstein2010}), which approximates the
posterior distribution by employing a specific Markov chain to sample
from it. This requires us to be able to point-wise compute unbiased
estimates of the posterior. In PMH, these are provided by the particle
filter \citep{GordonSalmondSmith1993,DoucetJohansen2011}, which is
another important and rather general statistical simulation algorithm.

The main aim and contribution of this tutorial is to give a gentle introduction (both in terms of methodology and software) to the PMH algorithm. We assume that the reader is familiar with traditional time series analysis and Kalman filtering at the level of \cite{BrockwellDavis2002}. Some familiarity with Monte Carlo approximations, importance sampling and Markov chain Monte Carlo (MCMC) at the level of \cite{Ross2012} is also beneficial.

The focus is on implementing the particle filter and PMH algorithms
step-by-step in the programming language \proglang{R} \citep{R}. We
employ \textit{literate programming}\footnote{The skeleton of the code
  is outlined as a collection of different variables denoted by
  \code{<variable>}. These parts of the code are then populated with
  content using a \proglang{C++} like syntax. Hence, the operator
  \code{=} results in the assignment of a piece of code to the
  relevant variable (overwriting what is already stored within it). On
  the other hand, \code{+=} denotes the operation of adding some
  additional code after the current code stored in the variable. The
  reader is strongly encouraged to print the source code from GitHub
  and read it in parallel with the tutorial to enable rapid learning
  of the material.} to build up the code using a top-down approach,
see Section~1.1 in \cite{PharrHumphreys2010}. The final implementation
is available as a \proglang{R} package \pkg{pmhtutorial}
\citep{pmhtutorial} distributed under the GPL-2 license via the
Comprehensive \proglang{R} Archive Network (CRAN) repository and
available at \url{https://CRAN.R-project.org/package=pmhtutorial}.
The \proglang{R} source code is also provided via GitHub with
corresponding versions for \proglang{MATLAB} \citep{MATLABR2017b} and
\proglang{Python} \citep{python} to enable learning by readers from
many different backgrounds, see Section~\ref{sec:conclusions:software}
for details.

There are a number of existing tutorials which relate to this one. A
thorough introduction to particle filtering is provided by
\cite{ArulampalamMaskellGordonClapp2002}, where a number of different
implementations are discussed and pseudo-code provided to facilitate
implementation. This is an excellent start for the reader who is
particularly interested in particle filtering. However, they do not
discuss nor implement PMH. The software \pkg{LibBi} and the connected
tutorial by \cite{Murray2015} includes high-performance
implementations and examples of particle filtering and PMH. The focus
of the present tutorial is a bit different as it tries to explain all
the steps of the algorithms and provides basic implementations in
several different programming languages. Finally,
\cite{AlaLuhtalaWhiteleyHeinePiche2016} presents an advanced tutorial
paper on twisted particle filtering. It also includes pseudo-code for
PMH but its focus is on particle filtering and does not discuss, e.g.,
how to tune the proposal in PMH. We return to discuss related work and
software throughout this tutorial in
Sections~\ref{sec:lgss:state:outlook},
\ref{sec:lgss:parameter:outlook}, \ref{sec:app:extensions} and
\ref{sec:software}.

The disposition of this tutorial is as follows. In
Section~\ref{sec:strategy}, we introduce the Bayesian parameter and
state inference problem in SSMs and discuss how and why the PMH
algorithm works. In Sections~\ref{sec:lgss:state} and
\ref{sec:lgss:parameter}, we implement the particle filter and PMH
algorithm for a toy model and study their properties as well as give
a small survey of results suitable for further study. In
Sections~\ref{sec:app:basic} and \ref{sec:app:extensions}, the PMH
implementation is adapted to solve a real-world problem from
finance. Furthermore, some more advanced modifications and
improvements to PMH are discussed and exemplified. Finally, we
conclude this tutorial by Section~\ref{sec:software} where related
software implementations are discussed.

\section{Overview of the PMH algorithm}
\label{sec:strategy}
In this section, we introduce the Bayesian inference problem in SSMs related to the parameters and the latent state. The PMH algorithm is then introduced to solve these intractable problems. The inclusion of the particle filter into PMH is discussed from two complementary perspectives: as an approximation of the acceptance probability and as auxiliary variables in an augmented posterior distribution.

\subsection{Bayesian inference in SSMs}
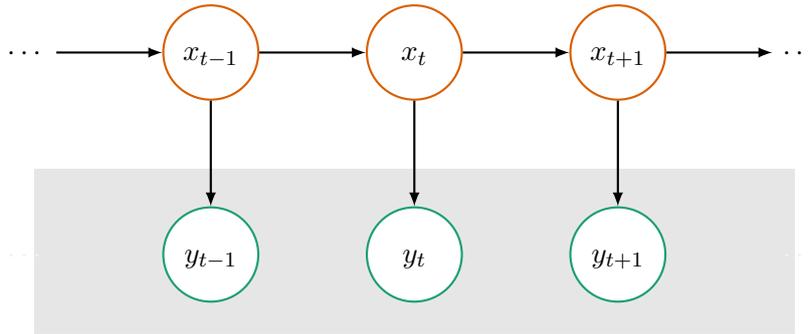
\begin{figure}[t!]
\centering
\tikzstyle{state}=[circle,
                                    thick,
                                    minimum size=1.25cm,
                                    draw=plotColors1,
                                    fill=white]

\tikzstyle{observation}=[circle,
                                                thick,
                                                minimum size=1.25cm,
                                                draw=plotColors0,
                                                fill=white]

\begin{tikzpicture}[>=latex,text height=1.5ex,text depth=0.25ex]
  \fill[bg] (-5,-2.4) rectangle (5,-0.2);
  \matrix[row sep=1.4cm,column sep=1.4cm] {
        \node (x_t-2)         {$\cdots$};           &
        \node (x_t-1) [state] {$x_{t-1}$}; &
        \node (x_t)   [state] {$x_t$};     &
        \node (x_t+1) [state] {$x_{t+1}$}; &
        \node (x_t+2)         {$\cdots$};
        \\
        \node (y_t-2)         {\textcolor{gray!10}{$\cdots$}};           &
        \node (y_t-1) [observation] {$y_{t-1}$}; &
        \node (y_t)   [observation] {$y_t$};     &
        \node (y_t+1) [observation] {$y_{t+1}$}; &
        \node (y_t+2)         {\textcolor{gray!10}{$\cdots$}};
        \\
    };

    \path[->] {
        (x_t-2) edge[thick] (x_t-1)
        (x_t-1) edge[thick] (x_t)
        (x_t)   edge[thick] (x_t+1)
        (x_t+1) edge[thick] (x_t+2)
        (x_t-1) edge[thick] (y_t-1)
        (x_t)   edge[thick] (y_t)
        (x_t+1) edge[thick] (y_t+1)
        }
        ;
\end{tikzpicture}
\caption{The structure of an SSM represented as a graphical model.}
\label{fig:cartoon-ssm}
\end{figure}

In Figure~\ref{fig:cartoon-ssm}, we present a graphical model of the
SSM with the latent states (top), the observations (bottom), but
without exogenous inputs\footnote{It is straightforward to modify the
  algorithms presented for vector valued quantities and also to
  include a known exogenous input $u_t$.}. The latent state $x_t$ only
depends on the previous state $x_{t-1}$ of the process as it is a
(first-order) Markov process. That is, all the information in the past
states $x_{0:t-1} \triangleq \{x_{s}\}_{s=0}^{t-1}$ is summarized in
the most recent state $x_{t-1}$. The observations $y_{1:t}$ are
conditionally independent as they only relate to each other through
the states.

We assume that the state and observations are real-valued, i.e.,
$y_t \in \mathcal{Y} \subseteq \mathbb{R}^{n_y}$ and
$x_t \in \mathcal{X} \subseteq \mathbb{R}^{n_x}$. The initial state
$x_0$ is distributed according to the density $\mu_{\theta}(x_0)$
parameterized by some unknown static parameters
$\theta \in \Theta \subset \mathbb{R}^p$. The latent state and the
observation processes are governed by the densities\footnote{In this
  tutorial, we assume that $x_t|x_{t-1}$ and $y_{t}|x_{t}$ can be
  modeled as continuous random variables with a density
  function. However, the algorithms discussed can be applied to
  deterministic states and discrete states/observations as well.}
$f_{\theta}(x_{t}|x_{t-1})$ and $g_{\theta}(y_t|x_t)$,
respectively. The density describing the state process gives the
probability that the next state is $x_t$ given the previous state
$x_{t-1}$. An analogous interpretation holds for the observation
process. With the notation in place, a general nonlinear SSM can be
expressed as
\begin{align}
	x_0           \sim \mu_{\theta}(x_0),
	\qquad
	x_{t}   | x_{t-1} \sim f_{\theta}( x_{t}   | x_{t-1} ),
	\qquad
	y_{t}   | x_t \sim g_{\theta}( y_{t}   | x_t ),
	\label{eq:ssm}
\end{align}
where no notational distinction is made between a random variable and its realization.

The main objective in Bayesian \textit{parameter inference} is to obtain an estimate of the parameters $\theta$ given the measurements $y_{1:T}$ by computing the parameter posterior (distribution) given by
\begin{align}
	\pi_{\theta}( \theta)
	\triangleq
	p( \theta | y_{1:T} )
	=
	\frac{
	p( \theta) p( y_{1:T} | \theta )
	}{
	\displaystyle
	\int_{\Theta} \! p( \theta' ) p( y_{1:T} | \theta' ) \dn \theta'
	}
	\triangleq
	\frac{\gamma_{\theta}(\theta)}{Z_{\theta}}
	,
	\label{eq:parameterposterior}
\end{align}
where $p( \theta )$ and
$p( y_{1:T} | \theta ) \triangleq p_{\theta}(y_{1:T})$ denote the
\textit{parameter prior (distribution)} and the \textit{data
  distribution}, respectively. The un-normalized posterior
distribution denoted by
$\gamma_{\theta}(\theta) = p( \theta) p( y_{1:T} | \theta )$ is an
important component in the construction of PMH. The denominator
$Z_{\theta}$ is usually referred to as the \textit{marginal
  likelihood} or the \textit{model evidence}.

The data distribution $p_{\theta}(y_{1:T})$ is intractable for most
interesting SSMs due to the fact that the state sequence is
unknown. The problem can be seen by studying the decomposition
\begin{align}
	p_{\theta}(y_{1:T})
	=
	p_{\theta}( y_{1} ) \prod_{t=2}^T p_{\theta}( y_{t} | y_{1:t-1} ),
	\label{eq:intro:likelihood-decomposition}
\end{align}
where the so-called \textit{predictive likelihood} is given by
\begin{align}
	p_{\theta}(y_{t}|y_{1:t-1})
	=
	\dint
	g_{\theta}(y_t|x_t)
	f_{\theta}( x_t|x_{t-1})
	\pi_{t-1}( x_{t-1} ) \dn x_t \dn x_{t-1}.
	\label{eq:predictive-likelihood}
\end{align}
The \textit{marginal filtering distribution} $\pi_{t-1}(x_{t-1}) \triangleq p_{\theta}(x_{t-1} | y_{1:t-1})$ can be obtained from the \textit{Bayesian filtering recursions} \citep{AndersonMoore2005}, which for a fixed $\theta$ are given by
\begin{align}
	\pi_{t}(  x_{t} )
	=
	\frac{
	g_{\theta}(y_t|x_t)
	}{
	p_{\theta}(y_{t} | y_{1:t-1} )
	}
	\,
	\int_{\mathcal{X}} \!
	f_{\theta}(x_{t} | x_{t-1} )
	\pi_{t-1}( x_{t-1} ) \dd x_{t-1}
	,
	\qquad
	\text{for } 0 < t \leq T,
	\label{eq:bayesianfilteringrecursions}
\end{align}
with $\pi_{0}( x_{0} ) = \mu_{\theta}( x_{0} )$ as the initialization. In theory, it is possible to construct a sequence of filtering distributions by iteratively applying \eqref{eq:bayesianfilteringrecursions}. However in practice, this strategy cannot be implemented for many models of interest as the marginalization over $x_{t-1}$ (expressed by the integral) cannot be carried out in closed form.

\subsection{Constructing the Markov chain}
\label{sec:strategy:markovchain}
The PMH algorithm offers an elegant solution to both of these intractabilities by leveraging statistical simulation. Firstly, a particle filter is employed to approximate $\pi_{t-1}(x_{t-1})$ and obtain point-wise unbiased estimates of the likelihood. Secondly, an MCMC algorithm \citep{RobertCasella2004} known as Metropolis-Hastings (MH) is employed to approximate $\pi_{\theta}(\theta)$ based on the likelihood estimator provided by the particle filter. PMH emerges as the combination of these two algorithms.

MCMC methods generate samples from a so-called \textit{target distribution} by constructing a specific Markov chain, which can be used to form an empirical approximation. In this tutorial, the target distribution will be either the parameter posterior $\pi_{\theta}(\theta)$ or the posterior of the parameters and states $\pi_{\theta,T}(\theta, x_{0:T}) = p(\theta, x_{0:T} | y_{1:T})$. We focus here on the former case and the latter follows analogously. Executing the PMH algorithm results in $K$ correlated samples $\theta^{(1:K)} \triangleq \{ \theta^{(k)} \}_{k=1}^K$ which can be used to construct a Monte Carlo approximation of $\pi_{\theta}(\theta)$. This \textit{empirical approximation} of the parameter posterior distribution is given by
\begin{align}
	\widehat{\pi}_{\theta}^K( \dn \theta )
	=
	\frac{1}{K}
	\sum_{k=1}^K
	\delta_{\theta^{(k)}}(\dn \theta),
	\label{eq:parameter-empricialdistribution}
\end{align}
which corresponds to a collection of Dirac delta functions
$\delta_{\theta'}(\dn \theta)$ located at $\theta=\theta'$ with equal
weights. In practice, histograms or kernel density estimators are
employed to visualize the estimate of the parameter posterior obtained
from \eqref{eq:parameter-empricialdistribution}.

The construction of the Markov chain within PMH amounts to carrying
out two steps to obtain one sample from $\pi_{\theta}(\theta)$. The
first step is to propose a so-called \textit{candidate parameter}
$\theta'$ from a \textit{proposal distribution}
$q(\theta'|\theta^{(k-1)})$ given the previous state of the Markov
chain denoted by $\theta^{(k-1)}$. The user is quite free to choose
this proposal but its support should cover the support of the target
distribution. The second step is to determine if the state of the
Markov chain should be changed to the candidate parameter $\theta'$ or
if it should remain in the previous state $\theta^{(k-1)}$. This decision
is stochastic and the candidate parameter is assigned as the next
state of the Markov chain, i.e.,
$\{ \theta^{(k)} \leftarrow \theta'\}$ with a certain
\textit{acceptance probability} $\alpha ( \theta', \theta^{(k-1)} )$
which is given by
\begin{align}
  \alpha \big( \theta', \theta^{(k-1)} \big)
  =  \min \Bigg\{ 1,
  \frac{ \pi_{\theta} \big( \theta' \big) }{ \pi_{\theta} \big( \theta^{(k-1)} \big) }
  \frac{ q \big( \theta^{(k-1)} \big| \theta' \big) }{ q \big( \theta' \big| \theta^{(k-1)} \big) }
  \Bigg\}
  =  \min \Bigg\{ 1,
  \frac{ \gamma_{\theta} \big( \theta' \big) }{ \gamma_{\theta} \big( \theta^{(k-1)} \big) }
  \frac{ q \big( \theta^{(k-1)} \big| \theta' \big) }{ q \big( \theta' \big| \theta^{(k-1)} \big) }
  \Bigg\},
  \label{eq:mhaprob}
\end{align}
where $Z_{\theta}$ cancels as it is independent of the current state
of the Markov chain. The stochastic behavior introduced by
\eqref{eq:mhaprob} facilitates exploration of (in theory) the entire
posterior and is also the necessary condition for the Markov chain to
actually have the target as its stationary distribution. The latter is
known as \textit{detailed balance} which ensures that the Markov chain
is \textit{reversible} and has the correct stationary
distribution. That is, to ensure that the samples generated by PMH
actually are from $\pi_{\theta}(\theta)$.

Hence, the stochastic acceptance decision can be seen as a correction of the Markov chain generated by the proposal distribution. The stationary distribution of the corrected Markov chain is the desired target distribution. In a sense this is similar to importance sampling, where samples from a proposal distribution are corrected by weighting to be approximately distributed according to the target distribution.

The intuition for the acceptance probability \eqref{eq:mhaprob}
(disregarding the influence of the proposal $q$) is that we always
accept a candidate parameter $\theta'$ if
$\pi_{\theta} \big( \theta' \big) > \pi_{\theta} \big( \theta^{(k-1)}
\big)$. That is if $\theta'$ increases the value of the target
compared with the previous state $\theta^{(k-1)}$. This results in a
mode-seeking behavior which is similar to that of an optimization
algorithm estimating the maximum of the posterior
distribution. However from \eqref{eq:mhaprob}, we also note that a
small decrease in the posterior value can be accepted to facilitate
exploration of the entire posterior. This is the main difference
between PMH and that of an optimization algorithm, where the former
focus on mapping the entire posterior whereas the latter only would
like to find the location of its mode.

\subsection{Approximating test functions}
\label{sec:strategy:testfunctions}
In Bayesian parameter inference, the interest often lies in computing the expectation of a so-called \textit{test function}, which is a well-behaved (integrable) function $\varphi: \Theta \rightarrow \mathbb{R}^{n_{\varphi}}$ mapping the parameters to a value on the real space. The expectation with respect to the parameter posterior is given by
\begin{align}
	\pi_{\theta}[\varphi]
	\triangleq
	\E \Big[ \varphi(\theta) \big| y_{1:T} \Big]
	=
	\int_{\Theta} \!
	\varphi(\theta)
	\pi_{\theta}( \theta ) \dd \theta,
	\label{eq:strategy:testfunction:theta}
\end{align}
where, e.g., choosing $\varphi(\theta)=\theta$ corresponds to
computing the mean of the parameter posterior. The expectation in
\eqref{eq:strategy:testfunction:theta} can be estimated via the
empirical approximation in \eqref{eq:parameter-empricialdistribution}
according to
\begin{align}
	\widehat{\pi}_{\theta}^K[\varphi]
	\triangleq
	\int_{\Theta} \!
	\varphi(\theta)
	\widehat{\pi}^K_{\theta} ( \dn \theta )
	=
	\frac{1}{K}
	\sum_{k=1}^K
	\int_{\Theta} \!
	\varphi(\theta)
	\delta_{\theta^{(k)}}(\dn \theta)
	=
	\frac{1}{K}
	\sum_{k=1}^K
	\varphi \big( \theta^{(k)} \big),
	\label{eq:testfunction-parameters}
\end{align}
where the last equality follows from the properties of the Dirac delta
function. This estimator is well-behaved and it is possible to
establish a law of large numbers (LLN) and a central limit theorem
(CLT), see \cite{MeynTweedie2009} for more information. From the LLN,
we know that the estimator is \textit{consistent} (and asymptotically
unbiased as $K \rightarrow \infty$). Moreover from the CLT, we know
that the error is approximately Gaussian with a variance decreasing by
$1/K$, which is the usual Monte Carlo rate. Note that the LLN usually
assumes independent samples but a similar result known as the
\textit{ergodic theorem} gives a similar result (under some
assumptions) even when $\theta^{(1:K)}$ are correlated\footnote{The
  theoretical details underpinning MH have been deliberately
  omitted. For the ergodic theorem to hold, the Markov chain needs to
  be \textit{ergodic}, which means that the Markov chain should be
  able to explore the entire state-space and not get stuck in certain
  parts of it. The interested reader can find more about the
  theoretical details in, e.g., \cite{Tierney1994} and
  \cite{MeynTweedie2009} for MH and \cite{AndrieuDoucetHolenstein2010}
  for PMH.}. However, the asymptotic variance is usually larger than
if the samples would be uncorrelated.

\subsection{Approximating the acceptance probability}
\label{sec:strategy:acceptanceprobability}
One major obstacle remains to be solved before the PMH algorithm can be implemented. The acceptance probability \eqref{eq:mhaprob} is intractable and cannot be computed as the likelihood is unknown. From above, we know that the particle filter can provide an unbiased point-wise estimator for the likelihood and therefore also the posterior $\pi_{\theta}(\theta)$. It turns out that the unbiasedness property is crucial for PMH to be able to provide a valid empirical approximation of $\pi_{\theta}(\theta)$. This approach is known as an \textit{exact approximation} due to that the likelihood is replaced by an approximation but the algorithm stills retains its validity, see \cite{AndrieuRoberts2009} for more information.

The particle filter generates a set of samples from $\pi_t( x_t)$ for each $t$, which can be used to create an empirical approximation. These samples are generated by sequentially applying importance sampling to approximate the solution to \eqref{eq:bayesianfilteringrecursions}. The approximation of the filtering distribution is then given by
\begin{align}
	\widehat{\pi}^N_{t}( \dn x_{t} )
	\triangleq
	\widehat{p}^N_{\theta}( \dn x_{t}|y_{1:t})
	=
	\sum_{i=1}^N
	\underbrace{
	\frac{
	v_t^{(i)}
	}{
	\sum_{j=1}^N
	v_t^{(j)}
	}}_{\triangleq w_t^{(i)}}
	\delta_{x_{t}^{(i)}}
	( \dn x_{t} ),
	\label{eq:lgss:state:empiricalfilteringdistribution}
\end{align}
where the \textit{particles} $x_t^{(i)}$ and their normalized weights
$w_t^{(i)}$ constitute the so-called \textit{particle system}
generated during a run of the particle filter. The un-normalized
weights are denoted by $v_t^{(i)}$. It is possible to prove that the
empirical approximation converges to the true distribution when
$N \rightarrow \infty$ under some regularity
assumptions\footnote{Again we have omitted many of the important
  theoretical details regarding the particle filter. For more
  information about these and many other important topics related to
  this algorithm, see, e.g.,\ \cite{CrisanDoucet2002} and
  \cite{DoucetJohansen2011}.}.

The likelihood can then be estimated via the decomposition
\eqref{eq:intro:likelihood-decomposition} by inserting the empirical
filtering distribution
\eqref{eq:lgss:state:empiricalfilteringdistribution} into
\eqref{eq:predictive-likelihood}. This results in the estimator
\begin{align}
	\widehat{p}_{\theta}^N(y_{1:T})
	&=
	\widehat{p}^N_{\theta}(y_1)
	\prod_{t=2}^T
	\widehat{p}^N_{\theta}( y_{t} | y_{1:t-1} )
	=
	\prod_{t=1}^{T}
	\left[
	\frac{1}{N}
	\sum_{i=1}^N v_t^{(i)}
	\right],
	\label{eq:likelihoodestimator}
\end{align}
where a point-wise estimate of the un-normalized parameter posterior
is given by
\begin{align}
	\widehat{\gamma}^N_{\theta}(\theta)
	=
	p(\theta)
	\widehat{p}_{\theta}^N(y_{1:T}).
	\label{eq:posteriorestimator}
\end{align}
Here, it is assumed that the prior can be computed point-wise by closed-form expressions. The acceptance probability \eqref{eq:mhaprob} can be approximated by plugging in \eqref{eq:posteriorestimator} resulting in
\begin{align}
  \alpha \big( \theta', \theta^{(k-1)} \big)
  &=
  \min \Bigg\{ 1,
  \frac{ \widehat{\gamma}^N_{\theta}(\theta') }{ \widehat{\gamma}^N_{\theta}(\theta^{(k-1)}) }
  \frac{ q \big( \theta^{(k-1)} \big| \theta' \big) }{ q \big( \theta' \big| \theta^{(k-1)} \big) }
  \Bigg\} \nonumber \\
  &=
  \min \Bigg\{ 1,
  \frac{ p(\theta') }
  { p(\theta^{(k-1)}) }
  \frac{ \widehat{p}^N_{\theta'}(y_{1:T}) }
  { \widehat{p}^N_{\theta^{(k-1)}}(y_{1:T}) }
  \frac{ q \big( \theta^{(k-1)} \big| \theta' \big) }
  { q \big( \theta' \big| \theta^{(k-1)} \big) }
  \Bigg\}.
  \label{eq:pmhaprob}
\end{align}
This approximation is only valid if $\widehat{p}_{\theta}^N(y_{1:T})$ is an unbiased estimator of the likelihood. Fortunately, this is the case when employing the particle filter as its likelihood estimator is unbiased \citep{DelMoral2013,PittSilvaGiordaniKohn2012} for any $N \geq 1$.

As in the parameter inference problem, interest often lies in computing an expectation of a well-behaved test function $\varphi: \mathcal{X} \rightarrow \mathbb{R}^{n_{\varphi}}$ with respect to $\pi_t(  x_t )$ given by
\begin{align*}
	\pi_t[\varphi]
	\triangleq
	\E \Big[ \varphi(x_t) \big| y_{1:t} \Big]
	=
	\int_{\mathcal{X}} \!
	\varphi(x_t)
	\pi_t ( x_t) \dd x_t,
\end{align*}
where again choosing the test function $\varphi(x_t)=x_t$ corresponds to computing the mean of the marginal filtering distribution. This expectation is intractable as $\pi_t ( x_t)$ is unknown. Again, we replace it with an estimate provided by an empirical approximation \eqref{eq:lgss:state:empiricalfilteringdistribution}, which results in
\begin{align}
	\widehat{\pi}^N_t[\varphi]
	\triangleq
	\int_{\mathcal{X}} \!
	\varphi(x_t)
	\widehat{\pi}_t^N( \dn x_t )
	=
	\sum_{i=1}^N
	w_t^{(i)}
	\int_{\mathcal{X}} \!
	\varphi(x_t)
	\delta_{x_{t}^{(i)}}
	( \dn x_t )
	=
	\sum_{i=1}^N
	w_t^{(i)}
	\varphi \Big( x_t^{(i)} \Big),
	\label{eq:lgss:state:estimatetestfunction}
\end{align}
for some $0 \leq t \leq T$ by the properties of the Dirac delta function. In the subsequent sections, we are especially interested in the estimator for the mean of the marginal filtering distribution,
\begin{align}
	\widehat{x}^N_t
	\triangleq
	\widehat{\pi}^N_t[x]
	=
	\sum_{i=1}^N
	w_t^{(i)}
	x_t^{(i)},
	\label{eq:lgss:state:filteredstate}
\end{align}
which is referred to as the \textit{filtered state estimate}.

Under some assumptions, the properties of $\widehat{\pi}^N_t[\varphi]$ are similar as for the estimator in the PMH algorithm, see \cite{CrisanDoucet2002} and \cite{DoucetJohansen2011} for more information. Hence, we have that the estimator is consistent (and asymptotically unbiased when $N \rightarrow \infty$) and the error is Gaussian with a variance decreasing by $1/N$.

\newpage

\subsection{Outlook: Pseudo-marginal algorithms}
\label{sec:strategy:pseudomarginal}
The viewpoint adopted in the previous discussion on PMH is that it is
an MH algorithm which employs a particle filter to approximate the
otherwise intractable acceptance probability. Another more general
viewpoint is to consider PMH as a \textit{pseudo-marginal algorithm}
\citep{AndrieuRoberts2009}. In this type of algorithm, some auxiliary
variables are introduced to facilitate computations but these are
marginalized away during a run of the algorithm.

This results in that the PMH algorithm can be seen as a standard MH algorithm operating on the non-standard extended space $\Theta \times \mathcal{U}$, where $\Theta$ and $\mathcal{U}$ denote the parameter space and the space of the auxiliary variables, respectively. The resulting \textit{extended target} is given by
\begin{align}
	\pi_{\theta, u}( \theta, u )
	=
	\widehat{\pi}_{\theta}^N ( \theta | u )
	\,
	m(u).
	\label{eq:pmmh:target}
\end{align}
Here, the parameter posterior is augmented with $u \in \mathcal{U}$ which denotes some multivariate random variables with density $m(u)$. From the discussion above, we know that $u$ can be used to construct a point-wise estimator of $\widehat{\pi}_{\theta}^N ( \theta | u )$ via the particle filter by \eqref{eq:posteriorestimator}.

The unbiasedness property of the likelihood estimator based on the particle filter gives
\begin{align}
  \E_{u}
	\Big[
	\widehat{\gamma}_{\theta}^N(\theta | u)
	\Big]
	=
	\int_{\mathcal{U}} \!
	\widehat{\gamma}_{\theta}^N(\theta | u)
	\,
	m(u)
	\dd u
	=
	\gamma_{\theta}( \theta ).
	\label{eq:pmmh:unbiasedest}
\end{align}
This means that the un-normalized parameter posterior
$\gamma_{\theta}(\theta)$ can be recovered by marginalizing over all
the auxiliary variables $u$. In the implementation, this results in
that the sampled values for $u$ simply can be disregarded. Hence, we
will not store them in the subsequent implementations but only keep
the samples of $\theta$ (and $x_t$).

We conclude by deriving the pseudo-marginal algorithm following from \eqref{eq:pmmh:target}. A proposal for $\theta$ and $u$ is selected with the form
\begin{align}
	q \left( \theta', u' \big| \theta^{(k-1)}, u^{(k-1)} \right)
	=
	q_{\theta} \left( \theta' \big| \theta^{(k-1)}, u^{(k-1)} \right)
	\,
	q_{u} \left( u' \big| u^{(k-1)} \right),
	\label{eq:pmh:propsal}
\end{align}
which is the product of the two proposals selected as
\begin{align}
	q_{\theta} \left( \theta' \big| \theta^{(k-1)}, u^{(k-1)} \right)
	=
	q_{\theta} \left( \theta' \big| \theta^{(k-1)} \right),
	\qquad
	q_{u} \left( u' \big| u^{(k-1)} \right) = m(u').
\end{align}
This corresponds to an independent proposal for $u$ and a proposal for $\theta$ that does not include $u$. Other options are a topic of current research, see Section~\ref{sec:lgss:parameter:outlook} for some references. The resulting acceptance probability from this choice of target and proposal is given by
\begin{align*}
	\alpha(\theta', u', \theta^{(k-1)}, u^{(k-1)})
	&=
	\min
	\left\{
	1,
	\frac{\pi_{\theta, u}( \theta', u' ) }
	{\pi_{\theta, u}( \theta^{(k-1)}, u^{(k-1)} )  }
	\frac{ q(\theta^{(k-1)}, u^{(k-1)} | \theta', u') }
	{ q(\theta', u' | \theta^{(k-1)}, u^{(k-1)} )}
	\right\} \\
	&=
	\min
	\left\{
	1,
	\frac{ \widehat{\gamma}_{\theta}^N(\theta' | u') }
	{ \widehat{\gamma}_{\theta}^N(\theta^{(k-1)} |u ^{(k-1)})  }
	\frac{ q_{\theta}(\theta^{(k-1)} | \theta') }
	{ q_{\theta}(\theta'| \theta^{(k-1)})}
	\right\},
\end{align*}
Note that the resulting acceptance probability is the same as in \eqref{eq:pmhaprob}.

These arguments provide a sketch of the proof that PMH generates samples from the target distribution and were first presented by \cite{FluryShephard2011}. For a more formal treatment and proof of the validity of PMH, see \cite{AndrieuRoberts2009} and \cite{AndrieuDoucetHolenstein2010}.

\section{Estimating the states in a linear Gaussian SSM}
\label{sec:lgss:state}
We are now ready to start implementing the PMH algorithm based on the material in Section~\ref{sec:strategy}. To simplify the presentation, this section discusses how to estimate the filtering distributions $\{\pi_{t}(x_t)\}_{t=0}^T$ for a linear Gaussian state-space (LGSS) model. These distributions can be used to compute $\widehat{x}^N_{0:T}$ and $\widehat{p}_{\theta}^N(y_{1:T})$, i.e., the estimates of the filtered state and the estimate of the likelihood, respectively. The parameter inference problem is treated in the subsequent section.

\begin{figure}[t!]
  \centering
  \includegraphics[width=0.95\textwidth]{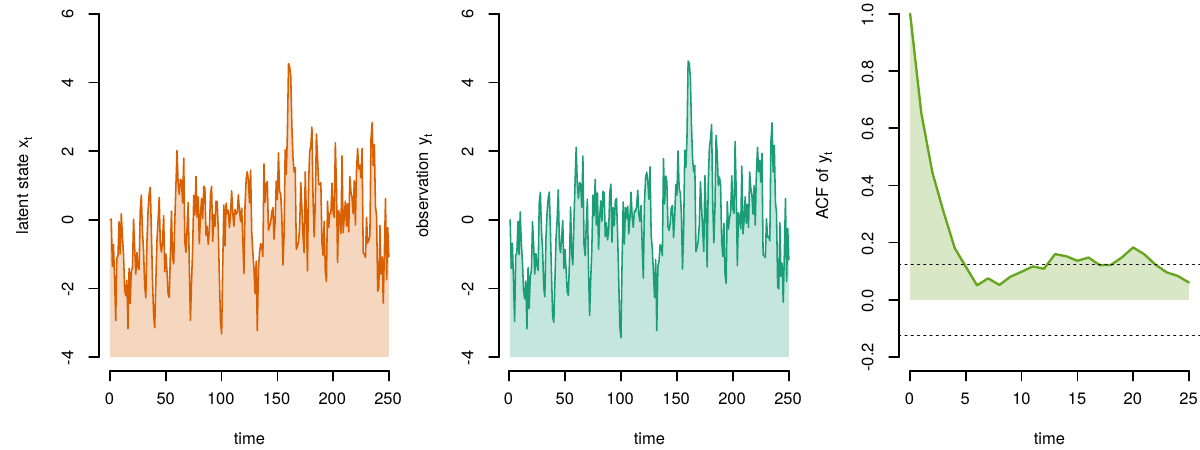}
  \caption{Simulated data from the LGSS model with latent state
    (orange), observations (green) and autocorrelation function
    (ACF) of the observations (light green).}
  \label{fig:lgss-data}
\end{figure}

The particular LGSS model considered is given by
\begin{align}
	x_0     \sim  \delta_{x_0}( x ),
	\qquad
	x_{t}   | x_{t-1} \sim \mathcal{N} \Big( x_{t} ; \phi x_{t-1}, \sigma_v^2 \Big),
	\qquad
	y_{t}   | x_t \sim \mathcal{N} \Big( y_{t}; x_t, \sigma_e^2 \Big),
	\label{eq:lgss:state:lgssmodel}
\end{align}
where parameters are denoted by $\theta=\{\phi,\sigma_v,\sigma_e\}$. $\phi \in (-1,1)$ determines the \textit{persistence} of the state, while $\sigma_v,\sigma_e \in \mathbb{R}_+$ denote the standard deviations of the state transition noise and the observation noise, respectively. The Gaussian density is denoted by $\mathcal{N}(x; \mu,\sigma^2)$ with mean $\mu$ and standard deviation $\sigma > 0$. Figure~\ref{fig:lgss-data} presents a simulated data record from the model with $T=250$ observations using $\theta=\{0.75,1.00,0.10\}$ and $x_0=0$. The complete code for the data generation is available in \code{generateData}.

\subsection{Implementing the particle filter}
\label{sec:lgss:state:implementation}
The complete source code for the implementation of the particle filter is available in the function \code{particleFilter}. Its code skeleton is given by:
\begin{Code}
particleFilter <- function(y, theta, noParticles, initialState) {
  <initialization>
  <initializeStates>
  for (t in 2:T) {
    <resampleParticles>
    <propagateParticles>
    <weightParticles>
    <normalizeWeights>
    <estimateLikelihood>
    <estimateState>
  }
  <returnEstimates>
}
\end{Code}
The function \code{particleFilter} has the inputs: \code{y} (vector
with $T+1$ observations), \code{theta} (parameters
$\{\phi,\sigma_v,\sigma_e\}$), \code{noParticles} and
\code{initialState}. The outputs will be the estimates of the filtered
states $\widehat{x}^N_{0:T}$ and the estimate of the likelihood
$\widehat{p}^N(\theta|y_{1:T})$. Note that \code{particleFilter}
iterates over $t = 2, \ldots, T$, which corresponds to time indices
$t = 1, \ldots, T-1$ in \eqref{eq:lgss:state:lgssmodel} as the
numbering of vector elements starts from $1$ in \proglang{R}. The
iteration terminates at time index $T-1$ as future observations
$y_{t+1}$ are required at each iteration. That is, we assume that the
new observation arrives to the algorithm between the propagation and
weighting steps. It therefore makes sense to use the information in
the weighting step if this is possible.

\code{<initialization>} The particle filter is initialized by allocating memory for the variables: \code{particles}, \code{ancestorIndices}, \code{weights} and \code{normalizedWeights}. These correspond to the particles, their ancestor, un-normalized weights and normalized weights, respectively. Moreover, the two variables \code{xHatFiltered} and \code{logLikelihood} are allocated to store $\widehat{x}^N_{0:T}$ and $\log \widehat{p}^N(\theta|y_{1:T})$, respectively. This is all implemented by:
\begin{Code}
<initialization> =
  T <- length(y) - 1

  particles <- matrix(0, nrow = noParticles, ncol = T + 1)
  ancestorIndices <- matrix(0, nrow = noParticles, ncol = T + 1)
  weights <- matrix(1, nrow = noParticles, ncol = T + 1)
  normalizedWeights <- matrix(0, nrow = noParticles, ncol = T + 1)
  xHatFiltered <- matrix(0, nrow = T, ncol = 1)
  logLikelihood <- 0
\end{Code}
\code{<initializeStates>} For the LGSS model \eqref{eq:lgss:state:lgssmodel}, we have that $\mu_{\theta}(x_0)=\delta_0(x_0)$ so all the particles are initially set to $x^{(1:N)}_0=x_0=0$ and all weights to $w^{(1:N)}_0 = 1/N$ (as all particles are identical). This operation is implemented by:
\begin{Code}
<initializeStates> =
  ancestorIndices[, 1] <- 1:noParticles
  particles[, 1] <- initialState
  xHatFiltered[, 1] <- initialState
  normalizedWeights[, 1] = 1 / noParticles
\end{Code}
%
\code{<resampleParticles>} The particles in the filter corresponds to
a number of different hypotheses of what the value of the latent state
could be. The weights represents the probability that a given particle
has generated the observation under the model. In the resampling step,
this information is used to focus the computational effort of the
particle filter to the relevant part of the state-space, i.e., to
particles with a large weight.

This is done by randomly duplicating particles with large weights and
discarding particles with small weights, such that the total number of
particles always remains the same. Note that, the resampling step is
unbiased in the sense that the expected proportions of the resampled
particles are given by the particle weights\footnote{This unbiasedness
  property is crucial for the PMH algorithm to be valid. As a
  consequence, adaptive methods \citep[Section~3.5]{DoucetJohansen2011}
  that only resamples the particle system at certain iterations cannot
  be used within PMH.}. This step is important as the particle system
otherwise would consist of only a single unique particle after a few
iterations. This would result in a large variance in
\eqref{eq:lgss:state:estimatetestfunction}.

In our implementation, we make use of \textit{multinomial} resampling,
which is also known as a weighted bootstrap with replacement. The
output from this procedure are the so-called \textit{ancestor indices}
$a^{(i)}_t$ for each particle $i$, which can be interpreted as the
parent index of particle $i$ at time $t$. For each $i=1,\ldots,N$, the
ancestor index is sampled from the multinomial (categorical)
distribution with
\begin{align*}
	\Prob \Big [a^{(i)}_t = j \Big] = w_{t-1}^{(j)}, \qquad j=1,\ldots,N.
\end{align*}
The resampling step is implemented by a call to the function \code{sample} by:
\begin{Code}
<resampleParticles> =
  newAncestors <- sample(noParticles, replace = TRUE,
                         prob = normalizedWeights[, t - 1])
  ancestorIndices[, 1:(t - 1)] <- ancestorIndices[newAncestors, 1:(t - 1)]
  ancestorIndices[, t] <- newAncestors
\end{Code}
where the resulting ancestor indices $a^{(1:N)}_t$ are returned as \code{newAncestors} and stored in \code{ancestorIndices} for bookkeeping. Note that the particle lineage is also resampled at the same time as the particles. All this is done to keep track of the genealogy of the particle system over time. That is, the entire history of the particle system.

\code{<propagateParticles>} The hypotheses regarding the state are updated by propagating the particle system forward in time by using the model. This corresponds to sampling from the \textit{particle proposal distribution} to generate new particles $x_t^{(i)}$ by
\begin{align}
	x_t^{(i)} | x_{t-1}^{a_t^{(i)}}
	\sim
	p_{\theta} \Big( x_t^{(i)} \big| x_{t-1}^{a_t^{(i)}}, y_t \Big),
	\label{eq:lgss:state:proposal}
\end{align}
where information from the previous particle $x_{t-1}^{a_t^{(i)}}$ and the current measurement $y_t$ can be included. There are a few different choices for the particle proposal and the most common one is to make use of the model itself, i.e., $p_{\theta}(x_t | x_{t-1}, y_t) = f_{\theta}(x_t | x_{t-1})$. However for the LGSS model, an \textit{optimal proposal} can be derived using the properties of the Gaussian distribution resulting in
\begin{align}
	p_{\theta}^{\text{opt}} \Big( x_t^{(i)} \big| x_{t-1}^{a_t^{(i)}}, y_{t} \Big)
	&\propto
	g_{\theta} ( y_{t} \big| x_{t}) f_{\theta} \Big( x_{t} \big| x_{t-1}^{a_t^{(i)}} \Big)
	\nonumber \\
	&=
	\mathcal{N} \Big(
	x_t^{(i)};
	\sigma^2 \Big[ \sigma_e^{-2}  y_t + \sigma_v^{-2} \phi x_{t-1}^{a_t^{(i)}} \Big]
	,
	\sigma^2
	\Big), \label{eq:lgss:state:propagationLGSS}
\end{align}
with $\sigma^{-2} = \sigma_v^{-2} + \sigma_e^{-2}$. This particle
proposal is optimal in the sense that it minimizes the variance of the
incremental particle weights at the current time
step\footnote{However, it is unclear exactly how this influences the
  entire particle system, i.e., if this is the globally optimal
  choice.}. For most other SSMs, the optimal proposal is intractable
and the state transition model is used instead. The propagation step
is implemented by:
\begin{Code}
<propagateParticles> =
  part1 <- (sigmav^(-2) + sigmae^(-2))^(-1)
  part2 <- sigmae^(-2) * y[t]
  part2 <- part2 + sigmav^(-2) * phi * particles[newAncestors, t - 1]
  particles[, t] <- part1 * part2 + rnorm(noParticles, 0, sqrt(part1))
\end{Code}
From \eqref{eq:lgss:state:propagationLGSS}, the ratio between the noise variances is seen to determine the shape of the proposal. Essentially, there are two different extreme situations (i) $\sigma_e^2 / \sigma_v^2 \ll 1$ and (ii) $\sigma_e^2 / \sigma_v^2 \gg 1$. In the first extreme (i), the location of the proposal is essentially governed by $y_t$ and the scale is mainly determined by $\sigma_e$. This corresponds to essentially simulating particles from $g_{\theta}(y_t|x_t)$. When $\sigma_e$ is small this usually allows for running the particle filter using only a small number of particles. In the second extreme (ii), the proposal essentially simulates from $f_{\theta}(x_t|x_{t-1})$ and does not take the observation into account. In summary, the performance and characteristics of the optimal proposal are therefore related to the noise variances of the model and their relative sizes.

\code{<weightParticles>} and \code{<normalizeWeights>} In these steps, the weights required for the resampling step are computed. The weights can be computed using different so-called \textit{weighting functions} and a standard choice is the observation model, i.e., $v_t(x_t) = g_{\theta}(y_t | x_t)$. However for the LGSS model, we can instead derive an \textit{optimal weighting} function by again applying the properties of the Gaussian distribution to obtain
\begin{align*}
	v^{(i)}_t
	\triangleq
	p(y_{t+1}|x_t^{(i)})
	=
	\dint g_{\theta} \Big( y_{t+1} \big| x_{t+1} \Big) f_{\theta} \Big( x_{t+1} \big| x_{t}^{(i)} \Big) \dd x_{t+1}
	= \mathcal{N} \Big( y_{t+1}; \phi x_{t}^{(i)}, \sigma^2_v + \sigma_e^2 \Big).
\end{align*}
Remember that in this implementation, the new observation $y_{t+1}$ is introduced in the particle filter between the propagation and the weighting steps for the first time. This is the reason for the slightly complicated choice of time indices. However, the inclusion of the information in $y_{t+1}$ in this step is beneficial as it improves the accuracy of the filter. That is, we keep particles that after propagation are likely to result in the observation $y_{t+1}$.

In many situations, the resulting weights are small and it is therefore beneficial to work with shifted log-weights to avoid problems with numerical precision. This is done by applying the transformation
\begin{align*}
	\widetilde{v}_t^{(i)} = \log v_t^{(i)} - v_{\max}, \qquad \text{for } i=1,\ldots,N,
\end{align*}
where $v_{\max}$ denotes the largest element in $\log v^{(1:N)}_t$. The weights are then normalized (ensuring that they sum to one) by
\begin{align}
	w_t^{(i)}
	=
	\frac{
	\exp \Big( \widetilde{v}_t^{(i)} \Big)
	}{
	\sum_{j=1}^N
	\exp
	\Big( \widetilde{v}_t^{(j)} \Big)
	}
	=
	\frac{
	\exp (- v_{\max} ) \exp \Big( \log v_t^{(i)} \Big)
	}{
	\exp (- v_{\max} ) \sum_{j=1}^N \exp \Big( \log v_t^{(j)} \Big)
	}
	=
	\frac{
	v_t^{(i)}
	}{
	\sum_{j=1}^N v_t^{(j)}
	},
	\label{eq:lgss:state:weights}
\end{align}
where the shifts $-v_{\max}$ cancel and do not affect the relative sizes of the weights. Hence, the first and third expressions in \eqref{eq:lgss:state:weights} are equivalent but the first expression enjoys better numerical precision. The computation of the weights is implemented by:
\begin{Code}
<weightParticles> =
  yhatMean <- phi * particles[, t]
  yhatVariance <- sqrt(sigmae^2 + sigmav^2)
  weights[, t] <- dnorm(y[t + 1], yhatMean, yhatVariance, log = TRUE)

<normalizeWeights> =
  maxWeight <- max(weights[, t])
  weights[, t] <- exp(weights[, t] - maxWeight)
  sumWeights <- sum(weights[, t])
  normalizedWeights[, t] <- weights[, t] / sumWeights
\end{Code}
We remind the reader that we compare \code{y[t+1]} and
\code{particles[, t]} due to the convention for indexing arrays in \proglang{R},
which corresponds to $y_{t}$ and $x_{t-1}^{(1:N)}$ in the model,
respectively. This is also the reason for the comment regarding the
for-loop in \code{particleFilter}. We now see that the weights depend
on the next observation and that is why the loop needs to run to $T-1$
and not $T$.

\code{<estimateLikelihood>}
The likelihood $p(\theta|y_{1:t})$ can be estimated by inserting the un-normalized weights into \eqref{eq:likelihoodestimator}. However, it is beneficial to instead estimate the log-likelihood to enjoy better numerical precision as the likelihood often is small. This results in rewriting \eqref{eq:likelihoodestimator} to obtain the recursion
\begin{align*}
	\log \widehat{p}_{\theta}^N(y_{1:t})
	=
	\log \widehat{p}_{\theta}^N(y_{1:t-1})
	+
	\left\{
	v_{\text{max}}
	+
	\log
	\left[
	\sum_{i=1}^N
	\exp
	\left(
	\tilde{v}_t^{(i)}
	\right)
	\right]
	- \log N
	\right\},
\end{align*}
where the log-shifted weights are used to increase the numerical precision. Note that the shift by $v_{\text{max}}$ needs to be compensated for in the estimation of the log-likelihood. This recursion is implemented by:
\begin{Code}
<estimateLikelihood> =
  predictiveLikelihood <- maxWeight + log(sumWeights) - log(noParticles)
  logLikelihood <- logLikelihood + predictiveLikelihood
\end{Code}
The estimate of the posterior distribution follows from inserting the estimate of the log-likelihood into \eqref{eq:posteriorestimator}.

\code{<estimateState>} The latent state $x_t$ at time $t$ can be estimated by \eqref{eq:lgss:state:filteredstate} given the observations $y_{1:(t+1)}$, which corresponds to the estimator
\begin{align}
	\widehat{x}^N_{t} = \frac{1}{N} \sum_{i=1}^N x_t^{(i)},
	\label{eq:lgss:state:meanfilterstateestimator}
\end{align}
which is implemented by:
\begin{Code}
<estimateState> =
  xHatFiltered[t] <- mean(particles[, t])
\end{Code}
Note that this corresponds to the weights $w_t^{(i)} = 1/N$ which
would correspond to that all particles are identical according to the
expression in \eqref{eq:lgss:state:weights}. However, this is not the
case in general and the estimator
\eqref{eq:lgss:state:meanfilterstateestimator} is the result of making
use of the optimal choices in the propagation and weighing steps of
the particle filter. This choice corresponds to a particular type of
algorithm known as the fully-adapted particle filter (faPF;
\citealp{PittShephard1999}), see Section~\ref{sec:lgss:state:outlook}
for more details. In a faPF, we make use of separate weights in the
resampling step and when constructing the empirical approximation of
$\pi_t( x_t)$. In Section~\ref{sec:app:basic}, we introduce another type
of particle filter with another estimator for $\widehat{x}^N_t$, which can be
used for most SSMs.

\code{<returnEstimates>} The outputs from \code{particleFilter} are returned by:
\begin{Code}
<returnEstimates> =
  list(xHatFiltered = xHatFiltered, logLikelihood = logLikelihood)
\end{Code}
%
\subsection{Numerical illustration}
\label{sec:lgss:state:example}
\begin{figure}[h!]
  \centering
  \includegraphics[width=0.99\textwidth]{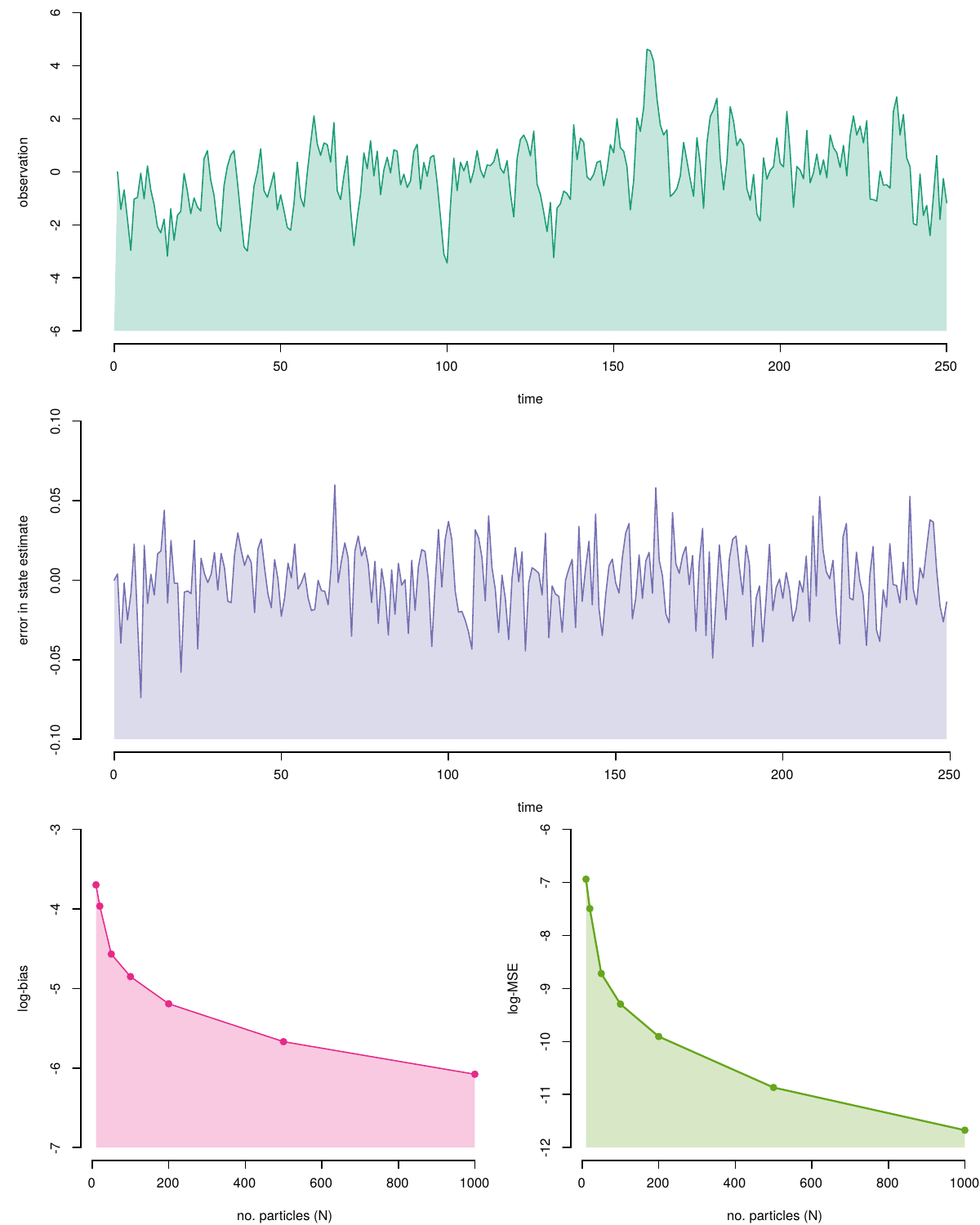}
  \caption{Top and middle: A simulated set of observations (top) from
    the LGSS model and the error in the latent state estimate (middle)
    using a particle filter with $N=20$. Bottom: the estimated
    log-bias (left) and log-MSE (right) for the particle filter when
    varying the number of particles $N$.}
	\label{fig:example1-lgss}
\end{figure}

We make use of the implementation of \code{particleFilter} to estimate
the filtered state and to investigate the properties of this estimate
for finite $N$. The complete implementation is found in the
script/function \code{example1_lgss}. We use the data presented in
Figure~\ref{fig:lgss-data}, which was generated in the beginning of
this section. The estimates from the particle filter are compared with
the corresponding estimates from the Kalman filter. The latter follows
from using the properties of the Gaussian distribution to solve
\eqref{eq:bayesianfilteringrecursions} exactly, which is only possible
for an LGSS model.

In the middle of Figure~\ref{fig:example1-lgss}, we present the difference between the optimal state estimate from the Kalman filter and the estimate from the particle filter using $N=20$ particles. Two alternative measures of accuracy are the bias (absolute error) and the mean square error (MSE) of the state estimate. These are computed according to
\begin{align*}
    \text{Bias} \big( \widehat{x}^N_{t} \big)
    =
    \frac{1}{T}
    \sum_{t=1}^T
	\big| \widehat{x}^N_{t} - \widehat{x}_{t} \big|,
    \qquad
    \text{MSE} \big( \widehat{x}^N_{t} \big)
    =
    \frac{1}{T}
    \sum_{t=1}^T
    \big( \widehat{x}^N_{t} - \widehat{x}_{t} \big)^2,
\end{align*}
where $\widehat{x}_{t}$ denotes the optimal state estimate obtained by the Kalman filter. In Table~\ref{tbl:smc-mse-versus-N} and in the bottom part of Figure~\ref{fig:example1-lgss}, we present the logarithm of the bias and the MSE for different values of $N$. We note that the bias and the MSE decrease rapidly when increasing $N$. Hence, we conclude that for this model $N$ does not need to be large for $\widehat{x}^N_t$ to be accurate.

\begin{table}[t!]
\centering
\begin{tabular}{lccccccc}
\hline
Number of particles ($N$) & 10 & 20 & 50 & 100 & 200 & 500 & 1000 \\
\hline
log-bias & $-$3.70 & $-$3.96 & $-$4.57 & $-$4.85 & $-$5.19 & \phantom{0}$-$5.67 & \phantom{0}$-$6.08 \\
log-MSE  & $-$6.94 & $-$7.49 & $-$8.72 & $-$9.29 & $-$9.91 & $-$10.87 & $-$11.67 \\
\hline
\end{tabular}
\caption{The log-bias and the log-MSE of the filtered states for varying $N$.}
\label{tbl:smc-mse-versus-N}
\end{table}

\subsection{Outlook: Design choices and generalizations}
\label{sec:lgss:state:outlook}
We make a short de-tour before the tutorial continues with the
implementation of the PMH algorithm. In this section, we provide some
references to important improvements and further studies of the
particle filter. Note that the scientific literature concerning the
particle filter is vast and this is only a small and naturally biased
selection of topics and references. More comprehensive surveys and
tutorials of particle filtering are given by
\cite{DoucetJohansen2011}, \cite{CappeGodsillMoulines2007},
\cite{ArulampalamMaskellGordonClapp2002} and
\cite{AlaLuhtalaWhiteleyHeinePiche2016} which also provide
pseudo-code. The alterations discussed within these papers can easily
be incorporated into the implementation developed within this section.

In Section~\ref{sec:lgss:state:implementation}, we employed the faPF
\citep{PittShephard1999} to estimate the filtered state and
likelihood. This algorithm made use of the optimal choices for the
particle proposal and weighting function, which corresponds to being
able to sample from $p(x_{t+1}|x_{t},y_{t+1})$ and to evaluate
$p(y_{t+1}|x_{t})$. This is not possible for many SSMs. One
possibility is to create Gaussian approximations of the required
quantities, see \cite{DoucetdeFreitasGordon2001} and
\cite{PittSilvaGiordaniKohn2012}. However, these methods rely on
quasi-Newton optimization that can be computationally prohibitive if
$N$ is large. See \cite{ArulampalamMaskellGordonClapp2002} for a
discussion and for pseudo-code. Another possibility is to make use of
a mixture of proposals and weighting functions in the particle filter
as described by \cite{KronanderSchon2014}. This type of filters are
based on multiple importance sampling, which is commonly used in,
e.g., computer graphics.

A different approach is to make use of an additional particle filter
to approximate the fully adapted proposal, resulting in a nested
construction where one particle filter is used within another particle
filter to construct a proposal distribution for that particle
filter. The resulting construction is referred to as nested sequential
Monte Carlo (SMC), see \cite{NaessethLindstenSchon2015} for
details. The nested SMC construction makes it possible to consider
state-spaces where $n_x$ is large, something that other types of
particle filter struggle with.

The bootstrap particle filter (bPF) is the default choice as it can be
employed for most SSMs. A drawback with the bPF is that it suffers
from poor statistical accuracy when the dimension of the state-space
$n_x$ grows beyond about $10$. In this setting, e.g., faPF or nested
SMC are required for state inference. However, there are some
approaches which possibly could improve the performance of the bPF in
some cases and should be considered for efficient
implementations. These includes: (i) parallel implementations, (ii)
better resampling schemes, (iii) make use of linear substructures of
the SSM and (iv) using quasi-Monte Carlo.

Parallel implementations of the particle filter (i) is a topic for
ongoing research but some encouraging results are reported by, e.g.,
\cite{LeeYauGilesDoucetHolmes2010} and \cite{LeeWhiteley2016}. In this
tutorial, we make use of multinomial resampling in the particle
filter. Alternative resampling schemes (ii) can be useful in
decreasing the variance of the estimates, see
\cite{HolSchonGustafsson2006} and \cite{DoucCappe2005} for some
comparisons. In general, systematic resampling is recommended for
particle filtering.

Another possible improvement is the combination of Kalman and particle
filtering (iii), which is possible if the model is conditionally
linear and Gaussian in some of its states. The idea is then to make
use of Kalman filtering to estimate these states and particle
filtering for the remaining states while keeping the linear ones fixed
to their Kalman estimates. These types of models are common in
engineering and Rao-Blackwellization schemes like this can lead to a
substantial decrease in variance. See, e.g.,
\cite{DoucetdeFreitasMurphyRussell2000}, \cite{ChenLiu2000} and
\cite{SchonGustafssonNordlund2005} for more information and
comparisons.

Quasi-Monte Carlo (iv) is based on so-called quasi-random numbers,
which are generated by deterministic recursions to better fill the
space compared with standard random numbers. These are useful in
standard Monte Carlo to decrease the variance in estimates. The use of
quasi-Monte Carlo in particle filtering is discussed by
\cite{GerberChopin2015}.

Finally, particle filtering is an instance of the SMC method
\citep{DelMoralDoucetJasra2006}, which represents a general class of
algorithms based on importance sampling. SMC can be employed for
inference in many statistical models and is a complement/alternative
to MCMC. A particularly interesting member is the SMC$^2$ algorithm
\citep{ChopinJacobPapaspiliopoulos2013,FulopLi2013}, which basically
is a two-level particle filter similar to nested SMC. The outer
particle filter maintains a particle system targeting
$\pi_{\theta}(\theta)$. The inner particle filter targets $\pi_t(x_t)$
and is run for each particle in the outer filter as this contains the
hypotheses of the value of $\theta$. The likelihood estimates from the
inner filter are used to compute the weights in the outer
filter. Hence, SMC$^2$ is an alternative to PMH for parameter
inference, see \cite{SvenssonSchon2016} for a comparison.

\section{Estimating the parameters in a linear Gaussian SSM}
\label{sec:lgss:parameter}
This tutorial now proceeds with the main event, where the PMH algorithm is implemented according to the outline provided in Section~\ref{sec:strategy}. The particle filter implementation from Section~\ref{sec:lgss:state:implementation} is employed to approximate the acceptance probability \eqref{eq:pmhaprob} in the PMH algorithm. We keep the LGSS model as a toy example to illustrate the implementation and provide an outlook of the PMH  literature at the end of this section.

A prior is required to be able to carry out Bayesian parameter inference. The objective in this section is to estimate the parameter posterior for $\theta=\phi$ while keeping $\{\sigma_v,\sigma_e\}$ fixed to their true values. Hence, we select the prior $p(\phi) = \mathcal{TN}_{(-1,1)}(\phi;0,0.5)$ to ensure that the system is stable (i.e., that the value of $x_t$ is bounded for every $t$). The truncated Gaussian distribution with mean $\mu$, standard deviation $\sigma$ in the interval $z \in [a,b]$ is defined by
\begin{align*}
	\mathcal{TN}_{[a,b]}(z;\mu,\sigma^2)
	=
	\mathbb{I}( a < z < b )
	\,
	\mathcal{N}(z; \mu, \sigma^2),
\end{align*}
where $\mathbb{I}(s)$ denotes the indicator function with value one if $s$ is true and zero otherwise.

\subsection{Implementing particle Metropolis-Hastings}
\label{sec:lgss:parameter:implementation}
The complete source code for the implementation of the PMH algorithm is available in the function \code{particleMetropolisHastings}. Its code skeleton is given by:
\begin{Code}
particleMetropolisHastings <- function(y, initialPhi, sigmav, sigmae,
  noParticles, initialState, noIterations, stepSize) {
  <initialization>
  for (k in 2:noIterations) {
    <proposeParameters>
    <computeAcceptProbability>
    <acceptRejectStep>
  }
  phi
}
\end{Code}
This function has inputs: \code{y} (vector with $T + 1$ observations),
\code{initialPhi} ($\phi^{(0)}$ the initial value for $\phi$),
\code{sigmav, sigmae} (parameters $\{\sigma_v,\sigma_e\}$),
\code{noParticles}, \code{initialState}, \code{noIterations} (no.\ PMH
iterations $K$) and \code{stepSize} (step size in the proposal). The
output from function \code{particleMetropolisHastings} is
$\phi^{(1:K)}$, i.e., the correlated samples approximately distributed
according to the parameter posterior $\pi_{\theta}$.

\code{<initialization>} We allocate some variables to store the
current state of the Markov chain \code{phi}, the proposed state
\code{phiProposed}, the current log-likelihood \code{logLikelihood}
and the proposed log-likelihood
\code{logLikelihoodProposed}. Furthermore, we allocate the binary
variable \code{proposedPhiAccepted} assuming the value $1$ if the
proposed parameter is accepted and $0$ otherwise. Finally, we run the
particle filter with the parameters $\{\phi^{(0)},\sigma_v,\sigma_e\}$
to estimate the initial likelihood. The initialization is implemented
by:
\begin{Code}
<initialization> =
  phi <- matrix(0, nrow = noIterations, ncol = 1)
  phiProposed <- matrix(0, nrow = noIterations, ncol = 1)
  logLikelihood <- matrix(0, nrow = noIterations, ncol = 1)
  logLikelihoodProposed <- matrix(0, nrow = noIterations, ncol = 1)
  proposedPhiAccepted <- matrix(0, nrow = noIterations, ncol = 1)

  phi[1] <- initialPhi
  theta <- c(phi[1], sigmav, sigmae)
  outputPF <- particleFilter(y, theta, noParticles, initialState)
  logLikelihood[1]<- outputPF$logLikelihood
\end{Code}
\code{<proposeParameters>} There are many choices for the proposal distribution as discussed in Section~\ref{sec:strategy}. In this tutorial, we make use of a \textit{Gaussian random walk} proposal given by
\begin{align}
	q \big( \theta' \big| \theta^{(k-1)} \big)
	=
	\mathcal{N} \big( \theta';\theta^{(k-1)}, \epsilon^2 \big),
    \label{eq:para:thproposal}
\end{align}
where $\epsilon > 0$ denotes the step size of the random walk, i.e., the standard deviation of the increment. The proposal step is implemented by:
\begin{Code}
<proposeParameters> =
  phiProposed[k] <- phi[k - 1] + stepSize * rnorm(1)
\end{Code}
\code{<computeAcceptProbability>} The acceptance probability \eqref{eq:pmhaprob} can be simplified since \eqref{eq:para:thproposal} is \textit{symmetric} in $\theta$, i.e.,
\begin{align*}
    q \big( \theta' \big| \theta^{(k-1)} \big)
    =
    q \big( \theta^{(k-1)} \big| \theta' \big),
\end{align*}
which results in that \eqref{eq:pmhaprob} can be expressed as
\begin{align*}
  \alpha(\theta', \theta^{(k-1)})
  =
  \min \Bigg\{ 1,
  \exp
  \left(
  \log \Bigg[
  \frac{ p(\theta') }{ p(\theta^{(k-1)}) }
  \Bigg]
  +
  \log
  \Bigg[
  \frac{ \widehat{p}^N_{\theta'}( y_{1:T} ) }{ \widehat{p}^N_{\theta^{(k-1)}}( y_{1:T} ) }
  \Bigg]
  \right)
  \Bigg\}.
\end{align*}
The log-likelihood and log-priors are used to avoid loss of numerical precision. The prior implies that $\alpha(\theta', \theta^{(k-1)}) =0$ when $|\phi|>1$. Therefore, the particle filter is not run in this case to save computations and the acceptance probability is set to zero to ensure that $\theta'$ is rejected. The computation of the acceptance probability is implemented by:
\begin{Code}
<computeAcceptProbability> =
  if (abs(phiProposed[k]) < 1.0) {
    theta <- c(phiProposed[k], sigmav, sigmae)
    outputPF <- particleFilter(y, theta, noParticles, initialState)
    logLikelihoodProposed[k] <- outputPF$logLikelihood
    }

  priorPart <- dnorm(phiProposed[k], log = TRUE)
  priorPart <- priorPart - dnorm(phi[k - 1], log = TRUE)
  likelihoodDifference <- logLikelihoodProposed[k] - logLikelihood[k - 1]
  acceptProbability <- exp(priorPart + likelihoodDifference)
  acceptProbability <- acceptProbability * (abs(phiProposed[k]) < 1.0)
\end{Code}
\code{<acceptRejectStep>} Finally, we need to take a decision for
accepting or rejecting the proposed parameter. This is done by
simulating a uniform random variable $\omega$ over $[0,1]$ by the
built-in \proglang{R} command \code{runif}. We accept $\theta'$ if
$\omega < \alpha \big( \theta',\theta^{(k-1)} \big)$ by storing it and
its corresponding log-likelihood as the current state of the Markov
chain. Otherwise, we keep the current values for the state and the
log-likelihood from the previous iteration. The accept/reject step is
implemented by:
\begin{Code}
<acceptRejectStep> =
  uniformRandomVariable <- runif(1)

  if (uniformRandomVariable < acceptProbability) {
    phi[k] <- phiProposed[k]
    logLikelihood[k] <- logLikelihoodProposed[k]
    proposedPhiAccepted[k] <- 1
  } else {
    phi[k] <- phi[k - 1]
    logLikelihood[k] <- logLikelihood[k - 1]
    proposedPhiAccepted[k] <- 0
  }
\end{Code}
%
\subsection{Numerical illustration}
\label{sec:lgss:parameter:example}
We make use of \code{particleMetropolisHastings} to estimate $\theta = \phi$ using the data in Figure~\ref{fig:lgss-data}. The complete implementation and code is available in the function/script \code{example2_lgss}. We initialize the Markov chain in $\theta_0=0.5$ and make use of $K=5,000$ iterations (\code{noIterations}) discarding the first $K_b=1,000$ iterations (\code{noBurnInIterations}) as \textit{burn-in}. That is, we only make use of the last $4,000$ samples to construct the empirical approximation of the parameter posterior to make certain that the Markov chain in fact has reached its stationary regime, see Section~\ref{sec:app:extensions} for more information.

\begin{figure}
    \centering
    \includegraphics[width=0.98\textwidth]{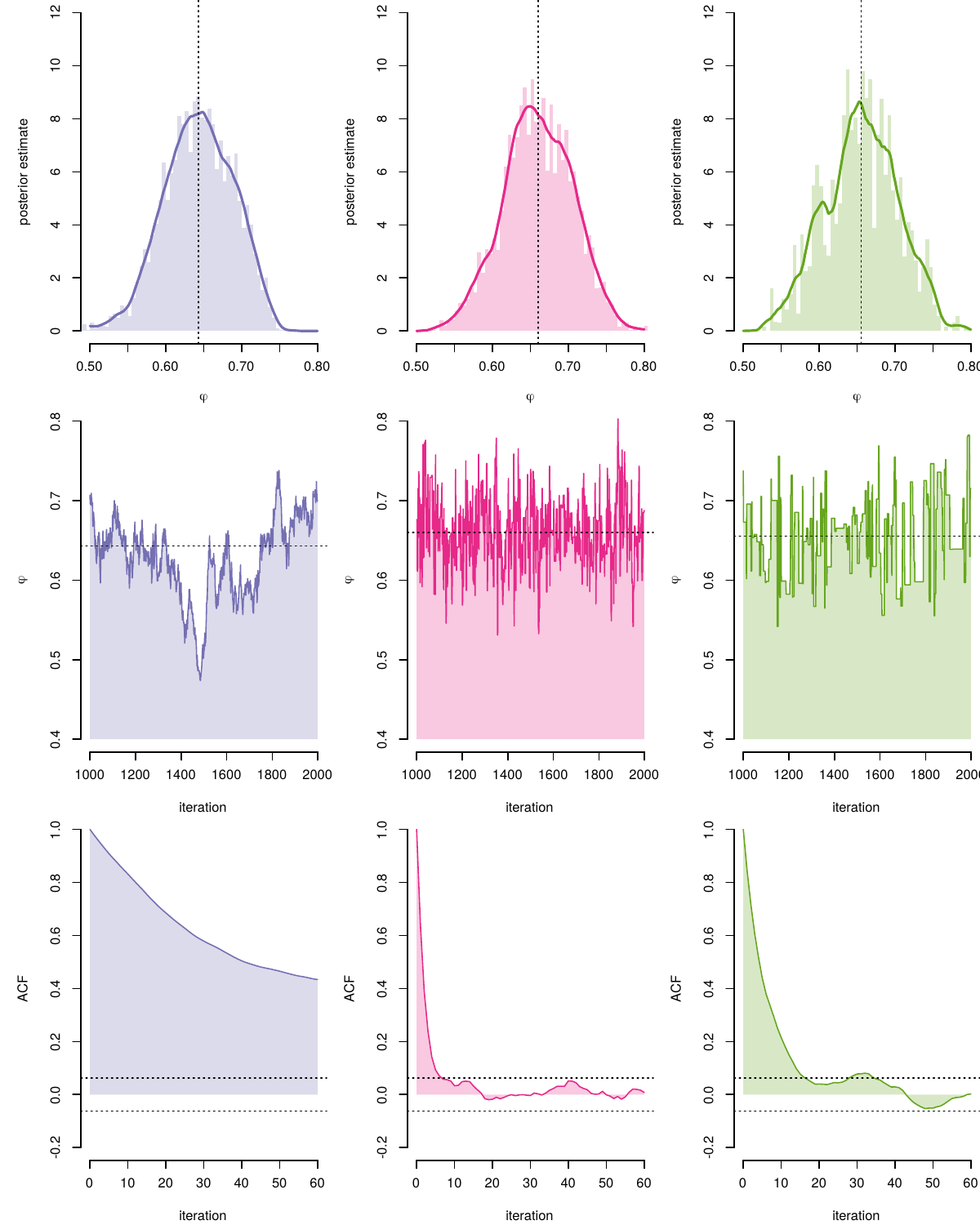}
    \caption{The estimate of $\pi_{\theta}[\phi]$ in the LGSS model
      using the PMH algorithm using three different step sizes:
      $\epsilon=0.01$ (left), $0.10$ (center) and $0.50$ (right). Top:
      the estimate of $\pi_{\theta}$ presented as a histogram and
      kernel density estimate (solid line). Middle: the state of the
      Markov chain at $1,000$ iterations after the burn-in. Bottom:
      the estimated ACF of the Markov chain. Dotted lines in the top
      and middle plots indicate the estimate of the posterior
      mean. The dotted lines in the bottom plot indicate the $95 \%$
      confidence intervals of the ACF coefficients.}
    \label{fig:lgss-parameter}
\end{figure}

In Figure~\ref{fig:lgss-parameter}, three runs of PMH are presented
using different step sizes $\epsilon=0.01$ (left), $0.10$ (center) and
$0.50$ (right). The resulting estimate of the posterior mean is
$\widehat{\phi} = 0.66$ for the case $\epsilon=0.10$. This is computed
by
\begin{CodeChunk}
\begin{CodeInput}
R> mean(phi[noBurnInIterations:noIterations])
\end{CodeInput}
\end{CodeChunk}
From the autocorrelation function (ACF), we see that the choice of
$\epsilon$ influences the correlation in the Markov chain. A good
choice of $\epsilon$ is therefore important to obtain an efficient
algorithm. We return to discussing this issue in
Section~\ref{sec:app:extensions:mixing}.

We note that the parameter estimate differs slightly from the true
value $0.75$ and that the uncertainty is rather large in the estimate
of the parameter posterior. This is due to the relatively small sample
size $T$ (and a finite $K$). From the asymptotic theory of the
Bayesian estimator, we know that the posterior mass tends to
concentrate around the true parameter as $T$ (and $K$) tends to
infinity.

We exemplify this in Table~\ref{tbl:pmh-posterior-versus-T} by
estimating the posterior mean and variance using the same setup when
$T$ increases. This small study supports that the true parameter is
indeed recovered by the posterior mean estimate in the limit. However,
the rate of this convergence is determined by the model and therefore
it is not possible to give any general guidelines for how large $T$
needs to be to achieve a certain accuracy.

\begin{table}[t!]
\centering
\begin{tabular}{lcccccc}
\hline
Number of observations ($T$)   & 10    & 20    & 50    & 100   & 200   & 500 \\
\hline
Estimated posterior mean     & 0.596 & 0.794 & 0.765 & 0.727 & 0.696 & 0.719\\
Estimated posterior variance & 0.040 & 0.013 & 0.009 & 0.006 & 0.003 & 0.001 \\
\hline
\end{tabular}
\caption{The estimated posterior mean and variance when varying $T$.}
\label{tbl:pmh-posterior-versus-T}
\end{table}

\subsection{Outlook: Generalizations}
\label{sec:lgss:parameter:outlook}
We devote this section to give some references for important
improvements and further studies of the PMH algorithm. As for the
particle filter, the scientific literature related to PMH is vast and
quickly growing. This is therefore only a small and biased selection
of recent developments. For a broader view of parameter inference in
SSM, see, e.g., the surveys by
\cite{KantasDoucetSinghMaciejowskiChopin2015} and
\cite{SchonLindstenDahlinWagbergNaessethSvenssonDai2015}.

As discussed in Section~\ref{sec:strategy}, the PMH algorithm is a
member of the family of exact approximation or pseudo-marginal
algorithms \citep{AndrieuRoberts2009}. Here, the particle filter is
used to estimate the target but it is also possible to make use of,
e.g., importance sampling for this end. PMH is also an instance of the
so-called particle MCMC (PMCMC; \citealp{AndrieuDoucetHolenstein2010})
algorithm, which also includes particle versions of Gibbs sampling
\citep{AndrieuDoucetHolenstein2010,LindstenJordanSchon2014}. PMCMC
algorithms are a topic of current research and much effort has been
given to improve their performance, see
Section~\ref{sec:app:extensions} for some examples of this effort.

In this tutorial, we make use of an independent proposal for $u$ as
discussed in Section~\ref{sec:strategy:pseudomarginal}. This
essentially means that all particle filters are independent. However,
intuitively there could be something to gain by correlating the
particle filters as the state estimates are often quite similar for
small changes in $\theta$. It turns out that correlating $u$ results
in a positive correlation in the estimates of the log-likelihood,
which decreases the variance of the estimate of the acceptance
probability. In practice, this means that $N$ does not need to scale
as rapidly with $T$ as for the case when $u$ are uncorrelated between
iterations. This is particularly useful when $T$ is large as it
decreases the computational cost of PMH. See
\cite{DahlinLindstenKronanderSchon2015},
\cite{ChoppalaGunawanChenTranKohn2016} and
\cite{DeligiannidisDoucetPitt2016} for more information and source
code.

\section{Application example: Estimating the volatility in OMXS30}
\label{sec:examples}
\label{sec:app:basic}
We continue with a concrete application of the PMH algorithm to infer
the parameters of a stochastic volatility (SV;
\citealp{HullWhite1987}) model. This is a nonlinear SSM with Gaussian
noise and inference in this type of model is an important problem as
the log-volatility (the latent state in this model) is useful for risk
management and to price various financial contracts. See, e.g.,
\cite{Tsay2005} and \cite{Hull2009} for more information. A particular
parameterization of the SV model is given by
\begin{subequations}
\begin{align}
    x_0           &\sim \mathcal{N} \bigg( x_0; \mu, \frac{ \sigma_v^2 }{ 1-\phi^2 } \bigg), \\
    x_{t+1} | x_t &\sim \mathcal{N} \Big( x_{t+1}; \mu + \phi ( x_t - \mu ), \sigma_v^2 \Big), \\
    y_{t}   | x_t &\sim \mathcal{N} \Big( y_{t}; 0, \exp(x_t) \Big),
\end{align}%
\label{eq:SVmodel}%
\end{subequations}%
\noindent where the parameters are denoted by $\theta = \{\mu,\phi,\sigma_v\}$. Here, $\mu \in \mathbb{R}$, $\phi \in [-1,1]$ and $\sigma_v \in \mathbb{R}_+$ denote the mean value, the persistence and standard deviation of the state process, respectively. Note that this model is quite similar to the LGSS model, but here the state $x_t$ scales the variance of the observation noise. Hence, we have Gaussian observations with zero mean and a state-dependent standard deviation given by $\exp(x_t/2)$.

\begin{figure}
    \centering
    \includegraphics[width=0.98\textwidth]{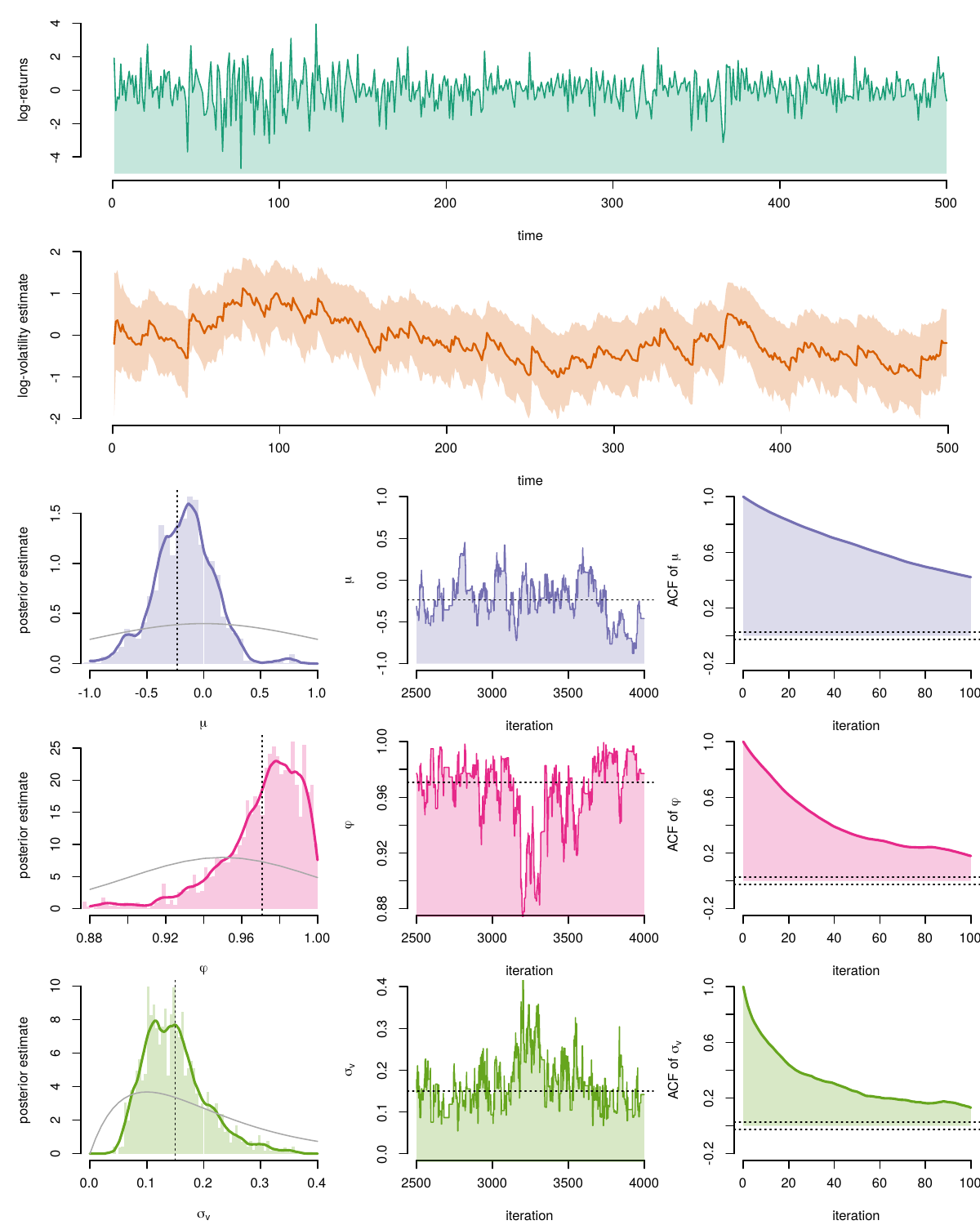}
    \caption{Top: the daily log-returns (dark green) and estimated
      log-volatility (orange) with $95\%$ confidence intervals of the
      NASDAQ OMXS30 index for the period between January 2, 2012 and
      January 2, 2014. Bottom: the posterior estimate (left), the
      trace of the Markov chain (middle) and the corresponding ACF
      (right) of $\mu$ (purple), $\phi$ (magenta) and $\sigma_v$
      (green) obtained from PMH. The dotted and solid gray lines in
      the left and middle plots indicate the parameter posterior mean
      and the parameter priors, respectively.}
    \label{fig:example3-sv}
\end{figure}

In econometrics, volatility is another word for standard deviation and therefore $x_t$ is known as the \textit{log-volatility}. The measurements in this model $y_t$ are so-called \textit{log-returns},
\begin{align*}
    y_t = 100 \, \log \left[ \frac{s_t}{s_{t-1}} \right] = 100\, \big[ \log(s_t) - \log(s_{t-1}) \big] ,
\end{align*}
where $s_t$ denotes the price of some financial asset (e.g., an index,
stock or commodity) at time $t$. Here, $\{s_t\}_{t=1}^T$ is the daily
closing prices of the NASDAQ OMXS30 index, i.e., a weighted average of
the $30$ most traded stocks at the Stockholm stock exchange. We
extract the data from Quandl\footnote{The data is available for
  download from: \url{https://www.quandl.com/data/NASDAQOMX/OMXS30}.}
for the period between January 2, 2012 and January 2, 2014. The
resulting log-returns are presented in the top part of
Figure~\ref{fig:example3-sv}. Note the time-varying persistent
volatility in the log-returns, i.e., periods of small and large
variations. This is known as the \textit{volatility clustering} effect
and is one of the features of real-world data that SV models aim to
capture. Looking at \eqref{eq:SVmodel}, this can be achieved when
$|\phi|$ is close to one and when $\sigma_v$ is small. As these
choices result in a first-order autoregressive process with a large
autocorrelation.

The objectives in this application are to estimate the parameters $\theta$ and the log-volatility $x_{0:T}$ from the observed data $y_{1:T}$. We can estimate both quantities using PMH as both samples from the posterior of the parameter and the state can be obtained at each iteration of the algorithm. To complete the SV model, we assume some priors for the parameters based on domain knowledge of usual ranges of the parameters, i.e.,
\begin{align*}
	p(\mu)=\mathcal{N}(\mu;0,1), \quad
	p(\phi)=\mathcal{TN}_{[-1,1]}(\phi;0.95,0.05^2), \quad
	p(\sigma_v) = \mathcal{G}(\sigma_v; 2, 10 ).
\end{align*}
Here, $\mathcal{G}(a,b)$ denotes a Gamma distribution with shape $a$ and scale $b$, i.e., expected value $a/b$.

\subsection{Modifying the implementation}
The implementation for the particle filter and PMH needs to be adapted
to this new model. We outline the necessary modifications by replacing
parts of the code in the skeleton for the two algorithms. The
resulting implementations and source codes are found in the functions
\code{particleFilterSVmodel} and
\code{particleMetropolisHastingsSVmodel}, respectively. In the
particle filter, we need to modify all steps except the resampling.

In the initialization, we need to simulate the initial particle system
from $\mu_{\theta}(x_0)$ by replacing:
\begin{Code}
<initializeStates> =
  ancestorIndices[, 1] <- 1:noParticles
  normalizedWeights[, 1] = 1 / noParticles
  particles[, 1] <- rnorm(noParticles, mu, sigmav / sqrt(1 - phi^2))
\end{Code}
Furthermore, we need to choose a (particle) proposal distribution and
the weighting function for the particle filter implementation. For the
SV model, we cannot compute a closed-form expression for the optimal
choices as for the LGSS model. Therefore, a bPF is employed which
corresponds to making use of the state dynamics as the proposal
\eqref{eq:lgss:state:proposal} given by
\begin{align*}
    x_t^{(i)} | x_{t-1}^{a_t^{(i)}}
    \sim
    f_{\theta} \Big( x_{t} \big| x_{t-1}^{a_t^{(i)}} \Big)
    = \mathcal{N} \bigg( x_t; \mu + \phi \Big( x_{t-1}^{a_t^{(i)}} - \mu \Big),\sigma_v^2 \bigg),
\end{align*}
and the observation model as the weighting function by
\begin{align*}
    W^{(i)}_t = g_{\theta} \Big( y_t \big| x_t^{(i)} \Big) = \mathcal{N} \Big( y_t; 0, \exp \big( x_t^{(i)} \big) \Big).
\end{align*}
These two choices result in that the estimator $\widehat{x}^N_t=\pi_t[x]$ is changed to
\begin{align*}
	\widehat{x}^N_{t} = \sum_{i=1}^N w^{(i)}_t x^{(i)}_t.
\end{align*}
These three alterations to the particle filter are implemented by replacing:
\begin{Code}
<propagateParticles> =
  part1 <- mu + phi * (particles[newAncestors, t - 1] - mu)
  particles[, t] <- part1 + rnorm(noParticles, 0, sigmav)

<weightParticles > =
  yhatMean <- 0
  yhatVariance <- exp(particles[, t] / 2)
  weights[, t] <- dnorm(y[t - 1], yhatMean, yhatVariance, log = TRUE)
  xHatFiltered[t] <- sum(normalizedWeights[, t] * particles[, t])
\end{Code}
We also need to generalize the PMH code to have more than one
parameter. This is straightforward and we refer the reader to the
source code for the necessary changes. The major change is to replace
the vectors \code{phi} and \code{phiProposed} with the matrices
\code{theta} and \code{thetaProposed}. That is, the state of the
Markov chain is now three-dimensional corresponding to the three
elements in $\theta$. Moreover, the parameter proposal is selected as
a multivariate Gaussian distribution centered around the previous
parameters $\theta^{(k-1)}$ with covariance matrix
$\Sigma = \mathsf{diag}( \epsilon )$, where $\epsilon$ denotes a
vector with three elements. This is implemented by:
\begin{Code}
<proposeParameters> =
  thetaProposed[k, ] <- rmvnorm(1, mean = theta[k - 1, ], sigma = stepSize)
\end{Code}
The computation of the acceptance probability is also altered to
include the new prior distributions. For this model, it is required
that $|\phi| < 1$ and $\sigma_v > 0$ to ensure that it is stable and
that the standard deviation is positive. This is implemented by:
\begin{Code}
<computeAcceptProbability> =
  if ((abs(thetaProposed[k, 2]) < 1.0) && (thetaProposed[k, 3] > 0.0)) {
    res <- particleFilterSVmodel(y, thetaProposed[k, ], noParticles)
    logLikelihoodProposed[k]  <- res$logLikelihood
    xHatFilteredProposed[k, ] <- res$xHatFiltered
  }

  priorMu <- dnorm(thetaProposed[k, 1], 0, 1, log = TRUE)
  priorMu <- priorMu - dnorm(theta[k - 1, 1], 0, 1, log = TRUE)
  priorPhi <- dnorm(thetaProposed[k, 2], 0.95, 0.05, log = TRUE)
  priorPhi <- priorPhi - dnorm(theta[k - 1, 2], 0.95, 0.05, log = TRUE)
  priorSigmaV <- dgamma(thetaProposed[k, 3], 2, 10, log = TRUE)
  priorSigmaV <- priorSigmaV - dgamma(theta[k - 1, 3], 2, 10, log = TRUE)
  prior <- priorMu + priorPhi + priorSigmaV

  likelihoodDifference <- logLikelihoodProposed[k] - logLikelihood[k - 1]
  acceptProbability <- exp(prior + likelihoodDifference)

  acceptProbability <- acceptProbability * (abs(thetaProposed[k, 2]) < 1.0)
  acceptProbability <- acceptProbability * (thetaProposed[k, 3] > 0.0)
\end{Code}
Furthermore, \code{<acceptRejectStep>} needs to be altered to take into account that \code{theta} is a vector, see the source code for the details.

\subsection{Estimating the log-volatility}
It is possible to compute an estimate of the log-volatility that takes into account the uncertainty in $\theta$ using the pseudo-marginal view of PMH. The particle filter is a deterministic algorithm given $u$. Hence, the random variables $u$ are equivalent to an estimate of the filtering distribution. The result is that it is quite simple to compute the state estimate by including it in the Markov chain generated by PMH.

This is done by modifying the code for the particle filter. After one
run of the algorithm, we sample a single trajectory by sampling a
particle at time $T$ with probability given by $w^{(1:N)}_T$. We then
follow the ancestor lineage back to $t=0$ and extract the
corresponding path in the state-space. This enables us to obtain a
better estimate of the log-volatility as both past and future
information are utilized. As a consequence, this often reduces the
variance of the estimate. This is done by using the stored resampled
ancestor indices and the sampled state trajectory by replacing:
\begin{Code}
<returnEstimates> =
  ancestorIndex  <- sample(noParticles, 1, prob = normalizedWeights[, T])
  sampledTrajectory <- cbind(ancestorIndices[ancestorIndex, ], 1:(T + 1))
  xHatFiltered <- particles[sampledTrajectory]
  list(xHatFiltered = xHatFiltered, logLikelihood = logLikelihood)
\end{Code}
in the function \code{particleFilterSVmodel}. The sampled state
trajectory in \code{xHatFiltered} is then treated in the same manner
as the candidate parameter in PMH, see the source code for details. As
a result, we can compute the posterior mean of the parameters and the
log-volatility and its corresponding standard deviation by:
\begin{CodeChunk}
\begin{CodeInput}
R> thetaStationary <- pmhOutput$theta[noBurnInIterations:noIterations, ]
R> thetaHatMean <- colMeans(thetaStationary)
R> thetaHatStandardDeviation <- apply(thetaStationary, 2, sd)
R> xHatStationary <- pmhOutput$xHatFiltered[2500:7500, ]
R> xhatFilteredMean <- colMeans(xHatStationary)
R> xhatFilteredStandardDeviation <- apply(xHatStationary, 2, sd)
\end{CodeInput}
\end{CodeChunk}
where \code{pmhOutput} denotes the output variable from \code{particleMetropolisHastingsSVmodel}.

\subsection{Numerical illustration}
\label{sec:app:basic:results}
We now make use of \code{particleMetropolisHastingsSVmodel} and \code{particleFilterSVmodel} to the SV model using the OMXS30 data introduced in the beginning of this section. The complete implementation is available in the function/code \code{example3_sv}. The resulting posterior estimates, traces and ACFs are presented in Figure~\ref{fig:example4-sv}. We see that the Markov chain and posterior are clearly concentrated around the posterior mean estimate $\widehat{\theta}=\{-0.23,0.97,0.15\}$ with standard deviations $\{0.37,0.02,0.06\}$. This confirms our belief that the log-volatility is a slowly varying process as $\phi$ is close to one and $\sigma_v$ is small. An estimate of the log-volatility given this parameter is also computed and presented in the second row of Figure~\ref{fig:example4-sv}. We note that the volatility is larger around $t=100$ and $t=370$, which corresponds well to what is seen in the log-returns.

\section{Selected advanced PMH topics}
\label{sec:app:extensions}
In this section, we outline a selected number of possible improvements
and \textit{best practices} for the implementation in
Section~\ref{sec:app:basic}. We discuss initialization, convergence
diagnostics and how to improve the so-called mixing of the Markov
chain. For the latter, we consider three different approaches: tuning
the random walk proposal, re-parameterizing the model and tuning the
number of particles. This section is concluded by a small survey of
more advanced proposal distributions and post-processing.

\subsection{Initialization}
It is important to initialize the Markov chain in areas of high
posterior probability mass to obtain an efficient algorithm. In
theory, we can initialize the chain anywhere in the parameter space
and still obtain a convergent Markov chain. However, in practice
numerical difficulties can occur when the parameter posterior assumes
small values or is relatively insensitive to changes in
$\theta$. Therefore it is advisable to try to obtain a rough estimate
of the parameters to initialize the chain closer to the posterior
mode.

In the LGSS model, we can make use of standard optimization algorithms in combination with a Kalman filter to estimate the mode of the posterior. However, this is not possible for general SSMs as only noisy estimates of the posterior can be obtained from the particle filter. Standard optimization methods can therefore get stuck in local minima or fail to converge entirely. One approach to mitigate this problem is the simultaneous perturbation and stochastic approximation (SPSA; \citealp{Spall1987}) algorithm, which can be applied in combination with the particle filter to estimate the posterior mode.

\subsection{Diagnosing convergence}
It is in general difficult to prove that the Markov chain has reached
its stationary regime and that the samples obtained are actually
samples from $\pi_{\theta}(\theta)$. A simple solution is to
initialize the algorithm at different points in the parameter space
and compare the resulting posterior estimates. This can give an
indication that the chain has reached its stationary regime if the
resulting estimates are similar.

Another alternative is to make use of the Kolmogorov-Smirnov (KS; \citealp{Massey1951}) test to establish that the posterior estimate does not change after the burn-in. This is done by dividing the samples $\{\theta^{(k)}\}_{k=1}^K$ into three partitions: the burn-in and two sets of equal number of samples from the stationary phase. As the KS test requires uncorrelated samples, \textit{thinning} can be applied to decrease the autocorrelation. The Markov chain is thinned by keeping every $l$th sample, where $l$ is selected such that the autocorrelation is negligible between two retained consecutive samples. A standard KS test is then conducted to conclude if the samples in the two thinned partitions are from the same (stationary) distribution or not.

Other methods for diagnosing convergence are discussed by \cite{RobertCasella2009} and \cite{GelmanCarlinSternDunsonVehtariRubin2013}.

\subsection{Improving mixing}
\label{sec:app:extensions:mixing}
Standard Monte Carlo methods assume that independent samples can be
obtained from the distribution of interest or from some instrumental
distribution. However, the samples obtained from PMH are correlated as
they are a realization generated from a Markov chain. Intuitively,
these correlated samples contain less information about the target
distribution than if the samples were independent. That is, most
samples are similar and the target is not fully explored if the
autocorrelation in the Markov chain is large. This autocorrelation is
also known as \textit{mixing}, which is a very important concept in
the MCMC literature. It is the key quantity for comparing the
efficiency of different MCMC algorithms.

It turns out that the mixing influences the performance of an MCMC
algorithm by scaling the asymptotic variance of the estimates. This is
apparent from the CLT governing $\widehat{\pi}_{\theta}^K[\varphi]$ in
\eqref{eq:testfunction-parameters} which under some
assumptions\footnote{These assumptions include that the Markov chain
  should reach its stationary regime quickly enough. This requirement
  is known as \textit{uniform ergodicity}, which means that the total
  variational norm between the distribution of the Markov chain and
  its stationary distribution should decrease with a geometric rate
  regardless of the initialization of the Markov chain, see, e.g.,
  \cite{Tierney1994} for more details.} has the form
\begin{align}
	\sqrt{K}
	\Big(
	\pi_{\theta}[\varphi] - \widehat{\pi}_{\theta}^K[\varphi]
	\Big)
	\stackrel{d}{\longrightarrow}
	\mathcal{N}
	\Big(
	0,
	\VAR_{\pi}[\varphi] \, \cdot \,
	\mathsf{IACT} \big( \theta^{(1:K)} \big)
	\Big), \qquad K \rightarrow \infty,
	\label{eq:MCMCCLT}
\end{align}
when
$\VAR_{\pi}[\varphi] = \pi_{\theta}[ ( \varphi - \pi_{\theta}[\varphi]
)^2 ] < \infty$. Here, $\mathsf{IACT} \big( \theta^{(1:K)} \big)$
denotes the integrated autocorrelation time (IACT) of the Markov
chain, which is computed as the area under the ACF. An interpretation
of the IACT is that it estimates the number of iterations of an MCMC
algorithm between obtaining two uncorrelated samples. Hence, the IACT
is one if the samples are independent, which is the case in, e.g.,
importance sampling. Minimizing the IACT is therefore the same as
maximizing the mixing in an MCMC algorithm.

In the right column of Figure~\ref{fig:example3-sv}, we present the ACF for the three parameters in the SV model. This information can be used to compute the corresponding IACTs by
\begin{align*}
    \mathsf{IACT}(\theta^{(1:K)})
    =
    1 + 2
    \sum_{\tau=1}^{\infty} \rho_{\tau} \big( \theta^{(1:K)} \big),
\end{align*}
where $\rho_{\tau}=\E[ (\theta^{(k)}-\pi_{\theta}[\theta])(\theta^{(k+\tau)}-\pi_{\theta}[\theta]) ] / \pi_{\theta}[(\theta-\pi_{\theta}[\theta])^2$ denotes the autocorrelation coefficient at lag $\tau$ for $\theta^{(1:K)}$. In practice, we cannot compute the IACT exactly as we do not know the autocorrelation coefficients for all possible lags. Also, it is difficult to estimate $\rho_{\tau}$ when $\tau$ is large and $K$ is rather small. A possible solution to this problem is to cut off the ACF estimate at some lag and only consider lags smaller than this limit $L$. In this tutorial, we make use of $L=100$ lags to estimate the IACT by
\begin{align*}
    \widehat{\mathsf{IACT}}(\theta^{(1:K)})
    =
    1 + 2
    \sum_{\tau=1}^{100} \widehat{\rho}^K_{\tau} \big( \theta^{(1:K)} \big),
\end{align*}
where $\widehat{\rho}^K_{\tau}=\mathsf{COR}(\varphi(\theta^{(k)}),\varphi(\theta^{(k+\tau)}))$ denotes the estimate of the lag-$\tau$ autocorrelation of $\varphi$. For the results presented in Figure~\ref{fig:example3-sv}, this corresponds to the IACT $\{ 135, 86, 66 \}$ for each of the three parameters.

\begin{figure}[t!]
    \centering
    \includegraphics[width=\textwidth]{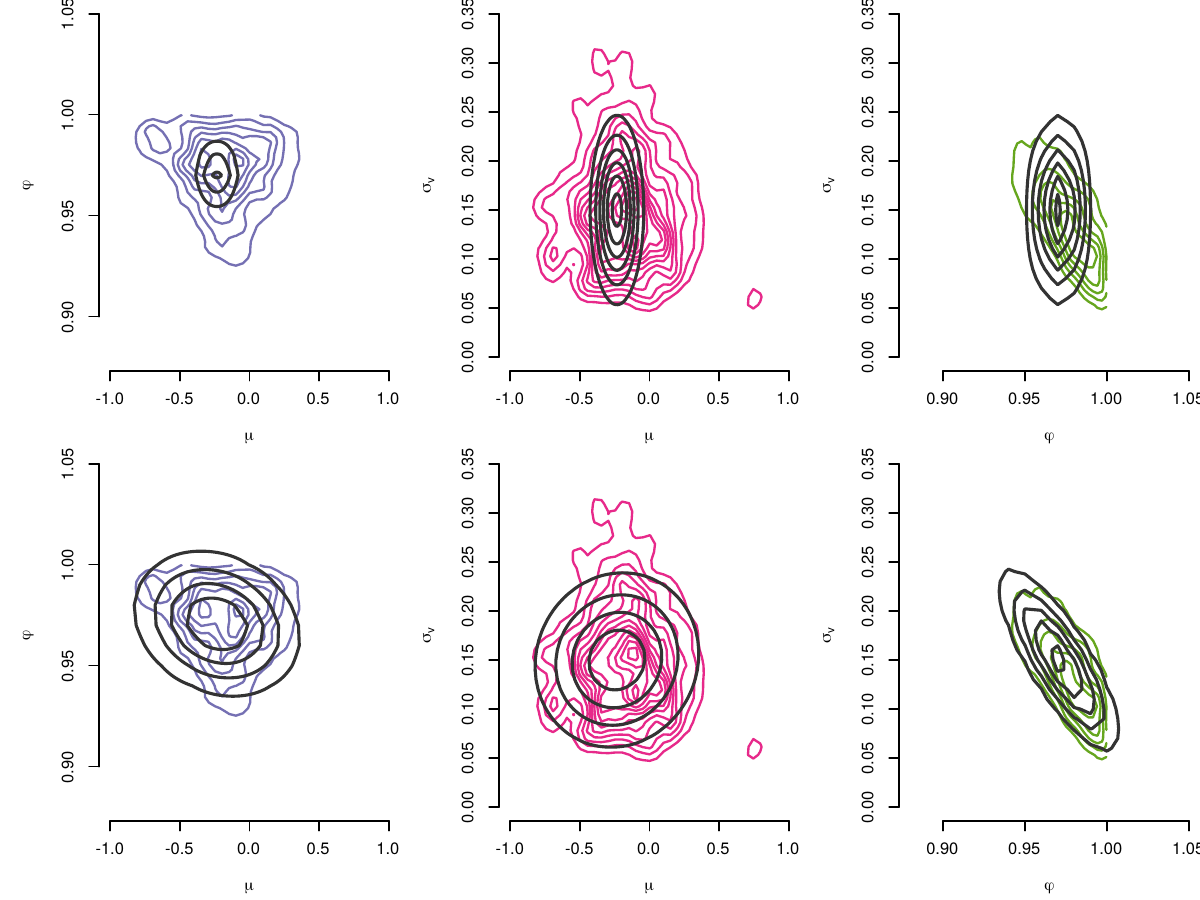}
    \caption{The estimated marginal posterior for $\mu$ and $\phi$
      (purple), $\mu$ and $\sigma_v$ (magenta) and $\phi$ and
      $\sigma_v$ (green) obtained from PMH. The dark contours
      (top/bottom) indicate the original proposal from
      Section~\ref{sec:app:basic} and the new (improved) proposal
      estimated from the pilot run, respectively. Both proposals are
      centered at the posterior mean estimate.}
    \label{fig:sv-2posteriors}
\end{figure}

\subsubsection{Tuning the proposal}
\label{sec:app:extensions:tuning}
The mixing can be improved by tuning the proposal distribution to
better fit the posterior distribution. This requires an estimate of
the posterior covariance, which acts like a \textit{pre-conditioning}
matrix $\mathcal{P}$. The covariance is quite simple to estimate from
a pilot run by
\begin{CodeChunk}
\begin{CodeInput}
R> var(thetaStationary)
\end{CodeInput}
\end{CodeChunk}
where \code{thetaStationary} denotes the trace of the Markov chain
generated by PMH after the burn-in has been discarded. For the problem
considered in Section~\ref{sec:app:basic}, this results in the
following pre-conditioning matrix
\begin{align*}
    \widehat{\mathcal{P}}
    =
    10^{-4}
    \begin{bmatrix}
        1371 & -16 &  15 \\
         -16 &   5 & -10 \\
          15 & -10 &  31 \\
    \end{bmatrix},
\end{align*}
which can be used to form a new improved proposal given by
\begin{align}
    q \left( \theta' \big| \theta^{(k-1)} \right)
	=
    \mathcal{N} \left(
    \theta';
    \theta^{(k-1)},
    \frac{2.562^2}{3} \widehat{\mathcal{P}}
    \right).
    \label{eq:app:sv2:improvedproposal}
\end{align}
This scaling of the covariance matrix was proposed by
\cite{SherlockThieryRobetsRosenthal2015} to minimize the IACT when
sampling from a Gaussian target distribution.

\begin{figure}
    \centering
    \includegraphics[width=0.98\textwidth]{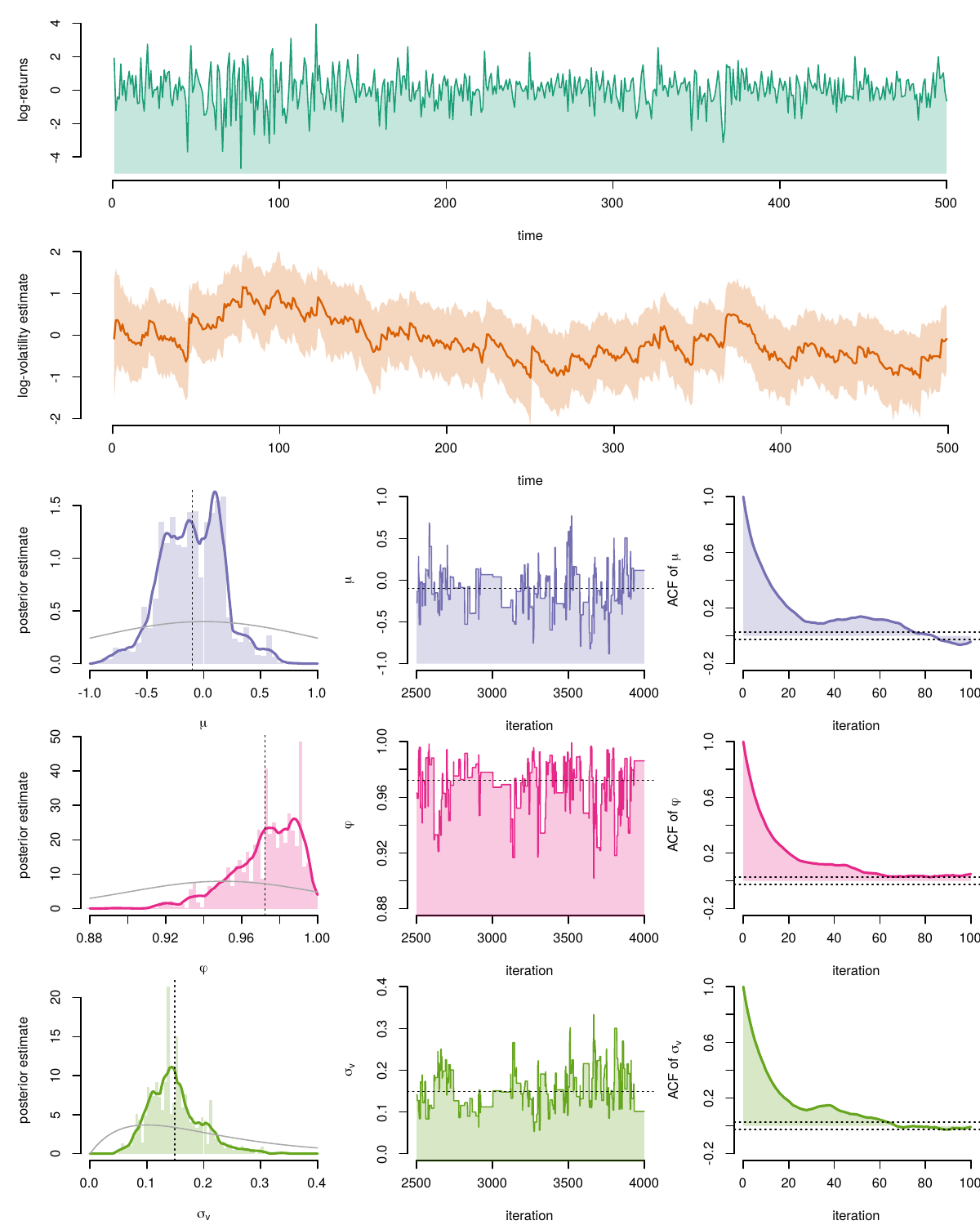}
    \caption{Top: the daily log-returns (dark green) and estimated
      log-volatility (orange) with $95\%$ confidence intervals of the
      NASDAQ OMXS30 index for the period between January 2, 2012 and
      January 2, 2014. Bottom: the posterior estimate (left), the
      trace of the Markov chain (middle) and the corresponding ACF
      (right) of $\mu$ (purple), $\phi$ (magenta) and $\sigma_v$
      (green) obtained from PMH. Dotted and solid gray lines in the
      left and middle plots indicate the parameter posterior mean and
      the parameter priors, respectively.}
    \label{fig:example4-sv}
\end{figure}

In Figure~\ref{fig:sv-2posteriors}, we present the marginal posteriors (in color) for each pair of parameters in \eqref{eq:SVmodel} together with the original (top) and the tuned proposals (bottom) from \eqref{eq:app:sv2:improvedproposal}. The tuned proposals fit the posteriors better and this is expected to improve the mixing of the Markov chain. In Figure~\ref{fig:example4-sv}, we present the results obtained by the implementation from Section~\ref{sec:app:basic} when using the tuned proposal instead. This code is available in the function \code{example4_sv}. The resulting IACT estimates are $\{ 32, 32, 28 \}$, which is a clear improvement in mixing for $\mu$ and $\sigma_v$ compared with the results obtained in Section~\ref{sec:app:basic:results}. The consequence is that $K$ (and therefore computational cost) can be cut by $135/32=4.2$ while retaining the same variance in the estimates. Note that the maximum IACTs are compared as these are the limiting factors.

We can also compare the support of the posterior estimates in Figures~\ref{fig:example3-sv} and \ref{fig:example4-sv}. According to \eqref{eq:MCMCCLT}, the improved mixing should decrease the asymptotic variance of the estimate. However, no such improvement can been seen in this case. This is probably due to that the variance in the posterior estimates largely results from the finite amount of information in the data.

\subsubsection{Re-parameterizing the model}
\label{sec:app:extensions:reparameterise}
Another approach to improve mixing is to re-parameterize the model to
obtain unconstrained parameters, which can assume any value on the
real line. In Section~\ref{sec:app:basic}, we constrained $\phi$ and
$\sigma_v$ to the regions $|\phi| < 1$ and $\sigma_v >0$ to obtain a
stable and valid SSM. This results in poor mixing if many candidate
parameters are proposed which violate these constraints and this
increases the autocorrelation. This problem can be mitigated by a
re-parameterization of the model given by
\begin{align}
    \phi = \tanh \big( \psi \big),
    \qquad
    \sigma_v = \exp \big( \varsigma \big),
    \label{eq:app:transformations}
\end{align}
such that $\psi,\varsigma \in \mathbb{R}$ are unconstrained parameters. Hence, the target of PMH is changed to the posterior of $\vartheta=\{\mu,\psi,\varsigma\}$. This transformation of parameters can be compensated for to still obtain samples from the posterior of $\theta$. This is done by taking the Jacobians of \eqref{eq:app:transformations} into account, which are given by
\begin{align}
    \frac{\partial}{\partial \psi} \tanh^{-1} ( \psi )
    = \frac{1}{1-\psi^2},
    \qquad
    \frac{\partial}{\partial \varsigma} \log ( \varsigma )
    = \varsigma^{-1}.
    \label{eq:pmh:sv:reparameterization:transformation}
\end{align}
The resulting acceptance probability is calculated according to
\begin{align}
  \alpha(\vartheta', \vartheta_{k-1})
  =
  \min \Bigg\{ 1,
  \frac{ p(\theta') }{ p(\theta^{(k-1)}) }
  \frac{ \widehat{p}^N_{\vartheta'}( y_{1:T} | u' ) }{ \widehat{p}^N_{\vartheta_{k-1}}( y_{1:T} | u_{k-1} ) }
  \left| \frac{ 1 - (\phi')^2 }{ 1 - \phi_{k-1}^2 } \right|
  \left| \frac{ \sigma'_v }{ \sigma_{v,k-1} } \right|
  \Bigg\},
  \label{eq:pmh:sv:reparameterization:aprob}
\end{align}
where the original parameters are used to compute the prior and to estimate the likelihood. Hence, the proposal operates on $\vartheta$ to propose a new candidate parameter $\vartheta'$ but \eqref{eq:pmh:sv:reparameterization:transformation} is applied to obtain $\theta'$, which is used to compute the acceptance probability \eqref{eq:pmh:sv:reparameterization:aprob}. After a run of this implementation, samples from the posterior of $\theta$ can be obtained by applying \eqref{eq:pmh:sv:reparameterization:transformation} to the samples from the posterior of $\vartheta$.

To implement this, we need to change the call to the particle filter
and the computation of the acceptance probability. The complete
implementation and code are available in the function
\code{example5_sv}. The resulting mean estimate of the parameter
posterior is $\{-0.16,0.96,0.17\}$ with standard deviation
$\{0.24,0.02,0.06\}$ and IACT $\{21, 29, 17\}$. Note that the maximum
IACT is even smaller than when tuning the proposal which resulted in a
speed-up of $4.7$ compared to the implementation in
Section~\ref{sec:app:basic}.

\subsubsection{Selecting the number of particles}
\label{sec:app:extensions:chossingN}
The particle filter plays an important role in the PMH algorithm and
greatly influences the mixing. If the log-likelihood estimates are
noisy (too small $N$ in the particle filter), the chain tends to get
stuck for several iterations and this leads to bad performance. This
is the result of the fact that sometimes
$p_{\theta'}(y_{1:T}) \ll \widehat{p}^N_{\theta'}(y_{1:T})$, which is
due to the stochasticity of the estimator. Thus, balancing $N$ and $K$
to obtain good performance at a reasonable computational cost is an
important problem. A small $N$ might result in that we need to take
$K$ large and vice verse. This problem is investigated by
\cite{PittSilvaGiordaniKohn2012} and \cite{DoucetPittKohn2012}. A
simple rule-of-thumb is to select $N$ such that the standard deviation
of the log-likelihood estimates is between $1.0$ and $1.7$. These
results are derived under certain strict assumptions that might not
hold in practice.

Here, we would like to investigate this rule-of-thumb in a practical
setting. We make use of the implementation when tuning the proposal
and re-run it for different choices of $N$. We also estimate the
standard deviation of the log-likelihood estimator from the particle
filter for each choice of $N$ using the estimate of the posterior mean
as the parameters and $1,000$ independent Monte Carlo runs. The
proposal distribution is tuned using a pilot run as before and the
Markov chain is initialized in the estimate of the posterior mean. The
IACT is computed using $L=100$ lags. Table~\ref{tbl:pmh-iact-versus-N}
presents some quantities related to the particle filter and PMH as $N$
is varied (keeping all other parameters fixed).

\begin{table}[t!]
\centering
\begin{tabular}{lcccccc}
\hline
Number of particles ($N$) &
50 & 100 & 200 & 300 & 400 & 500 \\
\hline
Standard deviation of $\widehat{p}^N_{\theta}(y_{1:T})$ &
2.6 & 1.8 & 1.2 & 1.0 & 0.9 & 0.7 \\
Acceptance probability (\%) &
6 & 14 & 22 & 28 & 29 & 33 \\
Maximum IACT &
165 & 139 & 128 & 92 & 114 & 97 \\
Time per PMH iteration (s) &
0.03 & 0.05 & 0.09 & 0.12 & 0.15 & 0.19 \\
Time per sample (s) &
5 & 7 & 11 & 11 & 17 & 18 \\
\hline
\end{tabular}
\caption{The standard deviation of the log-likelihood estimate, the acceptance probability, the maximum IACT for $\theta$, the computational time per iteration of PMH and a benchmark quantity (see main text) for varying $N$.}
\label{tbl:pmh-iact-versus-N}
\end{table}

The optimal choice for $N$ is between $100$ and $300$ depending on the
choice of proposal according to the rule-of-thumb proposed in \cite{DoucetPittKohn2012}. The objective for the user is often
to minimize the computational time of PMH to obtain a certain number
of uncorrelated samples from the posterior. Hence, a suitable
benchmark quantity is the maximum IACT multiplied with the
computational time for one iteration. Remember, that IACT is the
estimated number of iterations between two uncorrelated samples. The
benchmark quantity (last row of Table~\ref{tbl:pmh-iact-versus-N}) is
therefore an estimate of the computational time per uncorrelated
sample. The results are a bit inconclusive, which is the result of the
fact that the IACT is very challenging to estimate and thereby
noisy. However, the rule-of-thumb seems to be valid for this model and
setting $N$ to $100$ seems to be a good choice.

\subsection{Including geometric information}
In practical applications, we typically encounter performance issues
when the number of parameters $p$ grows beyond $5$ or as previously
discussed when the chain is initialized far from the areas of high
posterior probability. This problem occurs even if the rule-of-thumb
from \cite{SherlockThieryRobetsRosenthal2015} is used to tune the
proposal. The reason for this is that the problem lies within the use
of a random walk, which is known to poorly explore high-dimensional
spaces.

To mitigate this problem, it can be useful to take geometrical
information about the posterior into
account. \cite{GirolamiCalderhead2011} show how to make use of the
gradient and the Hessian of the log-posterior to guide the Markov
chain to areas of high posterior probability. In the paper, this is
motivated by diffusion processes on Riemann manifolds. However, a
perhaps simpler analogy is found in optimization, where we can make
use of noisy gradient ascent or Newton updates in the
proposal. Gradient information is useful to guide the Markov chain
towards the area of interest during the burn-in and also to keep it in
this area after the burn-in phase. The Hessian information can be used
to rescale the parameter posterior to make it closer to be isotropic,
which greatly simplifies sampling.

In \cite{DahlinLindstenSchon2015a}, the authors show how to make use
of this type of proposals in the PMH algorithm. We refer to the
proposal that makes use of only gradient information as PMH1 (for
first-order). The proposal that makes use of both gradient and Hessian
information is referred to as PMH2 (for second-order). The challenge
here is to obtain good estimates of the gradient and Hessian, which
both are analytically intractable for a nonlinear SSM. Furthermore,
the computation of these quantities usually requires the use of a
particle smoother with a high computational cost, which is prohibitive
inside the PMH algorithm. To mitigate this problem, they propose to
make use of the faster but more inaccurate fixed-lag particle smoother
and instead regularize the Hessian estimate when it is non-positive
definite.

In \cite{DahlinLindstenSchon2015c}, the authors describe a
quasi-Newton PMH2 proposal (qPMH2) based on a noisy quasi-Newton
update that does not require any Hessian information but constructs a
local approximation of the Hessian based on gradient information. PMH1
and similar algorithms are theoretically studied by
\cite{NemethSherlockFearnhead2016}, which offers a rule-of-thumb to
tune the step sizes based on an estimate of the posterior covariance.

\subsection{Control variates}
Control variates are a common and useful variance reduction technique
for standard Monte Carlo, which also can be applied to MCMC algorithms
such as PMH. Some interesting work along these lines for MH are found
in \cite{MiraSolgiImparato2013}, \cite{PapamarkouMiraGirolami2014},
\cite{MijatovicVogrinc2017} and \cite{DellaportasKontoyiannis2012},
which should be quite straightforward to implement for PMH. However
to the knowledge of the authors of this paper, control variates have
not yet been properly investigated for PMH but some encouraging
preliminary results are presented in Chapter~4.3 of \cite{Dahlin2016}
based on \cite{PapamarkouMiraGirolami2014}.

\section{Related software}
\label{sec:software}

There are a number of software packages related to the current
tutorial which implements (i) PMH and/or (ii) particle filtering and
SMC. These can be used to quickly carry out Bayesian parameter
inference in new SSMs or as building blocks for the user who would
like to create his/her own implementations of PMH.

The software \pkg{LibBi} \citep{Murray2015} provides a platform for
Bayesian inference in SSMs using both serial and parallel hardware
with a particular focus on high performance computing. It is written
in \proglang{C++} and allows the user to define new models using a
modeling language similar to \pkg{JAGS} \citep{JAGS} or \pkg{BUGS}
\citep{BUGS}. This enables the user to quickly solve the parameter
inference problem using PMH and SMC$^2$ in many different types of
SSMs. An example of using \pkg{LibBi} for inference in the SV model
introduced in Section~\ref{sec:app:basic} is offered via the
homepage\footnote{\url{http://libbi.org/}}, which is helpful in
learning how to use the software. Furthermore, interfaces for
\proglang{R} and \proglang{Octave} \citep{Octave} are available via
\pkg{RBi} via CRAN and \pkg{OctBi} via the \pkg{LibBi} homepage,
respectively. This software is suitable for the user who would like to
leverage the power of multi-core computing to carry out inference
quickly on large datasets.

Another alternative is \pkg{Biips}
\citep{TodeschiniCaronFuentesLegrandDelMoral2014}, which is
implemented in \proglang{C++} with interfaces to \proglang{R} and
\proglang{MATLAB} via the package \pkg{Rbiips} and the toolbox
\pkg{Matbiips}, respectively. The functionality of \pkg{Biips} is
similar to \pkg{LibBi} with the crucial difference on the focus on
parallel computations of the latter. Hence, \pkg{Biips} allows the
user to define a model using the \pkg{BUGS} language and carry out
inference using PMH. The
homepage\footnote{\url{https://biips.github.io/}} connected to
\pkg{Biips} contains some example code that implements the model from
Section~\ref{sec:app:basic}.

The \proglang{R} package \pkg{pomp} \citep{KingNguyenIonides2016}
includes an impressive setup of different inference methods for
SSMs. Among others, this includes particle filtering, PMH and also
approximate Bayesian computations (ABC;
\citealp{MarinPudloRobertRyder2012}). This software is a good choice
for the reader who uses \proglang{R} on a regular basis and would like
to explore some recent developments in statistical computing. It also
contains a number of pre-defined models for epidemiology. The model
specification in \pkg{pomp} is a bit more complicated compared with
\pkg{LibBi} and \pkg{Biips}, which could be a drawback for some users.

Probabilistic programming languages (PPLs) constitute a fairly recent
contribution to software for statistical modeling. They extend
graphical models with stochastic branches, such as loops and
recursions. Three popular PPLs which allow for carrying out inference
using PMH in SSMs are \pkg{Anglican}
\citep{TolpinVanDeMeentYangWood2016}, \pkg{Venture}
\citep{MansinghkaSelsamPerov2014} and \pkg{Probabilistic C}
\citep{PaigeWood2014}. \pkg{Anglican} is the most developed language
of these three and it runs as a standalone application. Implementation
of PMH in \pkg{Anglican} requires some additional coding compared
with, e.g., \pkg{LibBi}. However, it is a more general framework which
allows for inference for a larger class of models.

A \proglang{C++} template for particle filtering and more general SMC
algorithms is provided by \pkg{SMCTC} \citep{Johansen2009}. This
template is complemented with an interface to \proglang{R} via the
package \pkg{RCppSMC} \citep{EddelbuettelJohansen2017} and extended to
parallel computations in \proglang{C++} via the software \pkg{vSMC}
\citep{Zhou2015}. These templates require additional implementation by
the user to be able to carry out inference. They do not allow for easy
specification of new models or to carry out inference using, e.g., PMH
directly. However, they give the user full control of the particle
filter, which allows for adding extensions and tailoring the algorithm
to a specific problem.

A parallel implementation of SMC using {CUDA} is provided by
\cite{LeeYauGilesDoucetHolmes2010} and the source code is available
online\footnote{\url{http://www.oxford-man.ox.ac.uk/gpuss/cuda_mc.html}}. A
\proglang{MATLAB} toolbox for running SMC algorithms in parallel
called \pkg{DeCo} is developed by
\cite{CasarinGrassiRavazzoloVanDijk2015}. \pkg{DeCo} makes use of SMC
for combining densities connected to forecasts in economics, but could
potentially be generalized for other applications.

To conclude, we would like to briefly mention some additional
software. The \proglang{Python} library \pkg{pyParticleEst}
\citep{Nordh2017} provides functionality for state estimation using
different types of particle filters. It also includes parameter
estimation using maximum likelihood via the expectation-maximization
(EM; \citealt{DempsterLairdRubin1977,Mclachlan2007}) algorithm. MH is
implemented in \proglang{Ox} \citep{Ox} by \cite{Bos2011} to estimate
the states and parameters for the SV model in
Section~\ref{sec:app:basic}. In this setting, MH is applied to
estimate both the states and parameters, so no particle filter is
required. The \proglang{R} package \pkg{smfsb} \citep{Wilkinson2013a}
accompanying \cite{Wilkinson2011} contains demonstration code for
using PMH in a systems biology application.

The GitHub repository\footnote{\url{https://www.github.com/compops}.}
of the first author of this tutorial contains \proglang{Python} code
for PMH1, PMH2, qPMH2
\citep{DahlinLindstenSchon2015a,DahlinLindstenSchon2015c} and
pseudo-marginal MH with correlated random variables
\citep{DahlinLindstenKronanderSchon2015}. This code is quite similar
to the \proglang{Python} code supplied with this tutorial and is a
good starting point for the interested reader who would like to try
out more advanced implementations of PMH.

\section{Conclusions}
\label{sec:conclusions}
\label{sec:conclusions:software}
We have described the PMH algorithm for Bayesian parameter inference
in nonlinear SSMs. This includes the particle filter as it plays an
important role in PMH and provides an non-negative unbiased estimator
of the likelihood. Furthermore, we have applied PMH for inference in
an LGSS model and a SV model using both synthetic and real-world data.

We also identified and discussed a wide range of practical matters
related to PMH. This includes initialization and tuning of the
parameter proposal, which are important practical
problems. Furthermore, we have provided many references for the reader
who could like to continue his/her learning of particle filtering and
PMH. The implementation developed within this tutorial can be seen as
a compilation of minimal working examples. Hopefully, these code
snippets can be of use for the interested reader as a starting point
to develop his/her own implementations of the algorithms.

The implementations developed in this tutorial are available as the
package \pkg{pmhtutorial} from CRAN. Source code for the
implementation in \proglang{R} as well as similar code for
\proglang{MATLAB} and \proglang{Python} is available from GitHub at
\url{https://github.com/compops/pmh-tutorial}. See the
\code{README.md} files in the directory corresponding to each
programming language for specific comments and for dependencies.

\section*{Acknowledgments}
This work was supported by the projects: \textit{Probabilistic
  modeling of dynamical systems} (Contract number: 621-2013-5524),
CADICS, a Linnaeus Center, both funded by the Swedish Research Council
and \emph{ASSEMBLE} (Contract number: RIT15-0012) funded by the
Swedish Foundation for Strategic Research (SSF). The authors would
like to give a big thanks to Christian Andersson Naesseth, Wilfried
Bonou, Manon Kok, Joel Kronander, Fredrik Lindsten, Andreas Svensson
and Patricio Valenzuela for comments and suggestions that greatly
improved this tutorial.

The majority of the work was carried out while Johan Dahlin was
affiliated with the Division of Automatic Control, Link\"{o}ping University, Sweden.

\bibliography{ref}
\end{document}